   \documentstyle[12pt]{article}
\if@twoside\oddsidemargin25mm
\else\oddsidemargin25mm\evensidemargin25mm\marginparwidth25mm\fi%
\begin{document}
\newcommand{\ft}[2]{{\textstyle\frac{#1}{#2}}}
\newcommand{\QED}{{\hspace*{\fill}\rule{2mm}{2mm}\linebreak}}
\def\dop{{\rm d}\hskip -1pt}
\def\bfone{\relax{\rm 1\kern-.35em 1}}
\def\bfzero{\relax{\rm I\kern-.18em 0}}
\def\inbar{\vrule height1.5ex width.4pt depth0pt}
\def\IC{\relax\,\hbox{$\inbar\kern-.3em{\rm C}$}}
\def\ID{\relax{\rm I\kern-.18em D}}
\def\IF{\relax{\rm I\kern-.18em F}}
\def\IK{\relax{\rm I\kern-.18em K}}
\def\IH{\relax{\rm I\kern-.18em H}}
\def\II{\relax{\rm I\kern-.17em I}}
\def\IN{\relax{\rm I\kern-.18em N}}
\def\IP{\relax{\rm I\kern-.18em P}}
\def\IQ{\relax\,\hbox{$\inbar\kern-.3em{\rm Q}$}}
\def\IR{\relax{\rm I\kern-.18em R}}
\def\IG{\relax\,\hbox{$\inbar\kern-.3em{\rm G}$}}
\font\cmss=cmss10 \font\cmsss=cmss10 at 7pt
\def\ZZ{\relax\ifmmode\mathchoice
{\hbox{\cmss Z\kern-.4em Z}}{\hbox{\cmss Z\kern-.4em Z}}
{\lower.9pt\hbox{\cmsss Z\kern-.4em Z}}
{\lower1.2pt\hbox{\cmsss Z\kern-.4em Z}}\else{\cmss Z\kern-.4em
Z}\fi}
\def\a{\alpha} \def\b{\beta} \def\d{\delta}
\def\e{\epsilon} \def\c{\gamma}
\def\G{\Gamma} \def\l{\lambda}
\def\L{\Lambda} \def\s{\sigma}
+\def\cA{{\cal A}} \def\cB{{\cal B}}
\def\cC{{\cal C}} \def\cD{{\cal D}}
    \def\cF{{\cal F}} \def\cG{{\cal G}}
\def\cH{{\cal H}} \def\cI{{\cal I}}
\def\cJ{{\cal J}} \def\cK{{\cal K}}
\def\cL{{\cal L}} \def\cM{{\cal M}}
\def\cN{{\cal N}} \def\cO{{\cal O}}
\def\cP{{\cal P}} \def\cQ{{\cal Q}}
\def\cR{{\cal R}} \def\cV{{\cal V}}\def\cW{{\cal W}}
%
%
%
\def\crr{\crcr\noalign{\vskip {8.3333pt}}}
\def\tilde{\widetilde}
\def\bar{\overline}
\def\us#1{\underline{#1}}
\let\shat=\hat
\def\hat{\widehat}
\def\hyp{\vrule height 2.3pt width 2.5pt depth -1.5pt}
\def\square{\mbox{.08}{.08}}
\def\Coeff#1#2{{#1\over #2}}
\def\Coe#1.#2.{{#1\over #2}}
\def\coeff#1#2{\relax{\textstyle {#1 \over #2}}\displaystyle}
\def\coe#1.#2.{\relax{\textstyle {#1 \over #2}}\displaystyle}
\def\half{{1 \over 2}}
\def\shalf{\relax{\textstyle {1 \over 2}}\displaystyle}
\def\dag#1{#1\!\!\!/\,\,\,}
\def\to{\rightarrow}
\def\notin{\hbox{{$\in$}\kern-.51em\hbox{/}}}
\def\shdot{\!\cdot\!}
\def\ket#1{\,\big|\,#1\,\big>\,}
\def\bra#1{\,\big<\,#1\,\big|\,}
\def\equaltop#1{\mathrel{\mathop=^{#1}}}
\def\Trbel#1{\mathop{{\rm Tr}}_{#1}}
\def\inserteq#1{\noalign{\vskip-.2truecm\hbox{#1\hfil}
\vskip-.2cm}}
\def\attac#1{\Bigl\vert
{\phantom{X}\atop{{\rm\scriptstyle #1}}\phantom{X}}}
\def\exx#1{e^{{\displaystyle #1}}}
\def\del{\partial}
\def\delbar{\bar\partial}
\def\nex#1{$N\!=\!#1$}
\def\dex#1{$d\!=\!#1$}
\def\cex#1{$c\!=\!#1$}
\def\eg{{\it e.g.}} \def\ie{{\it i.e.}}
%
\def\cS{{\cal K}}
\def\IE{\relax{{\rm I\kern-.18em E}}}
\def\cE{{\cal E}}
\def\rt{{\cR^{(3)}}}
\def\IGam{\relax{{\rm I}\kern-.18em \Gamma}}
\def\IGa{\IA}
\def\LG{Lan\-dau-Ginz\-burg\ }
\def\cV{{\cal V}}
\def\Rt{{\cal R}^{(3)}}
\def\wabc{W_{abc}}
\def\WABC{W_{\a\b\c}}
\def\W{{\cal W}}
\def\tft#1{\langle\langle\,#1\,\rangle\rangle}
\def\IA{\relax{\hbox{{\rm A}\kern-.82em {\rm A}}}}
\let\picfuc=\fp
\def\hata{{\shat\a}}
\def\hatb{{\shat\b}}
\def\hatA{{\shat A}}
\def\hatB{{\shat B}}
\def\bv{{\bf V}}
\def\spg{special geometry}
\def\sc{SCFT}
\def\leel{low energy effective Lagrangian}
\def\pf{Picard--Fuchs}
\def\pfS{Picard--Fuchs system}
\def\el{effective Lagrangian}
\def\Fb{\overline{F}}
\def\nablab{\overline{\nabla}}
\def\Ub{\overline{U}}
\def\Db{\overline{D}}
\def\zb{\overline{z}}
\def\eb{\overline{e}}
\def\fb{\overline{f}}
\def\tb{\overline{t}}
\def\Xb{\overline{X}}
\def\Vb{\overline{V}}
\def\Cb{\overline{C}}
\def\Sb{\overline{S}}
\def\delb{\overline{\del}}
\def\Gammab{\overline{\Gamma}}
\def\Ab{\overline{A}}
\def\Anh{A^{\rm nh}}
\def\alphab{\bar{\alpha}}
\def\cy{Calabi--Yau}
\def\cabg{C_{\alpha\beta\gamma}}
\def\B{\Sigma}
\def\Bh{\hat \Sigma}
\def\Kh{\hat{K}}
\def\Knh{{\cal K}}
\def\A{\Lambda}
\def\Ah{\hat \Lambda}
\def\R{\hat{R}}
\def\V{{V}}
\def\T{T}
\def\Gammah{\hat{\Gamma}}
\def\twot{$(2,2)$}
\def\K{K\"ahler}
\def\rat{({\theta_2 \over \theta_1})}
\def\lv{{\bf \omega}}
\def\w{w}
\def\CP{C\!P}
\def\o#1#2{{{#1}\over{#2}}}
\newcommand{\be}{\begin{equation}}
\newcommand{\ee}{\end{equation}}
\newcommand{\ba}{\begin{eqnarray}}
\newcommand{\ea}{\end{eqnarray}}
\newtheorem{definizione}{Definition}[section]
\newcommand{\bd}{\begin{definizione}}
\newcommand{\ed}{\end{definizione}}
\newtheorem{teorema}{Theorem}[section]
\newcommand{\bth}{\begin{teorema}}
\newcommand{\eth}{\end{teorema}}
\newtheorem{lemma}{Lemma}[section]
\newcommand{\blem}{\begin{lemma}}
\newcommand{\elem}{\end{lemma}}
\newcommand{\brr}{\begin{array}}
\newcommand{\err}{\end{array}}
\newcommand{\nn}{\nonumber}
\newtheorem{corollario}{Corollary}[section]
\newcommand{\bcorol}{\begin{corollario}}
\newcommand{\ecorol}{\end{corollario}}
\def\twomat#1#2#3#4{\left(\begin{array}{cc}
 {#1}&{#2}\\ {#3}&{#4}\\
\end{array}
\right)}
\def\twovec#1#2{\left(\begin{array}{c}
{#1}\\ {#2}\\
\end{array}
\right)}
\begin{flushright}
CERN-TH/97-180 
\\
July 1997
\end{flushright}
\vskip 1cm
\begin{center}
{\LARGE {Flat Symplectic Bundles of $N$-Extended Supergravities,
 Central Charges and Black-Hole Entropy }}
\footnote{
\noindent
 Work supported in part by EEC under TMR contract ERBFMRX-CT96-0045
 (LNF Frascati, Politecnico di Torino and Univ. Genova) and by DOE grant
DE-FGO3-91ER40662.}\\
\vskip 1.5cm
  {\bf Laura Andrianopoli$^a$,
Riccardo D'Auria$^b$ and
Sergio Ferrara$^c$ } \\
\vskip 0.5cm
{\small
$^a$ Dipartimento di Fisica, Universit\'a di Genova, via Dodecaneso 33,
I-16146 Genova\\
and Istituto Nazionale di Fisica Nucleare (INFN) - Sezione di Torino, Italy\\
\vspace{6pt}
$^b$ Dipartimento di Fisica, Politecnico di Torino,\\
 Corso Duca degli Abruzzi 24, I-10129 Torino\\
and Istituto Nazionale di Fisica Nucleare (INFN) - Sezione di Torino, Italy\\
\vspace{6pt}
$^c$ CERN Theoretical Division, CH 1211 Geneva 23, Switzerland}
\end{center}
\begin{center}
{\bf Summary}
\footnote{
\noindent
Based on lectures given by S. Ferrara at the 5th Winter School on Mathematical
Physics held at the Asia Pacific Center for Theoretical Physics, Seul (Korea),
February 1997}
\end{center}
{\small{In these lectures we give a geometrical formulation of $N$-extended
supergravities which generalizes $N=2$ special geometry of $N=2$ theories.
\par
In all these theories duality symmetries are related to the notion of 
"flat symplectic bundles" and central charges may be defined as "sections" 
over these bundles.
Attractor points giving rise to "fixed scalars" of the horizon geometry and
Bekenstein-Hawking entropy formula for extremal black-holes are discussed in
some details.}} 


 \section{Introduction}
Recent developments on duality symmetries \cite{dual} in supersymmetric quantum
theories of fields and strings seem to indicate that the known
different string theories are different manifestations, in different
regions of the coupling constant space, of a unique more fundamental
theory that, depending on the regime and on the particular
compactification, may itself reveal extra (11 or 12) dimensions \cite{witten}, \cite{vafa}.
A basic aspect that allows a comparison of  different theories is their
number of supersymmetries and their spectrum of massless and massive
BPS states.
Indeed, to explore a theory in the nonperturbative regime, the
power of supersymmetry allows one to compute to a large extent
all dynamical details encoded in the low energy effective action of
a given formulation of the theory and to study the moduli (coupling
constants) dependence of the BPS states.
This latter property is important in order to study more dynamical
questions such as phase transitions in the moduli space \cite{huto}, \cite{wi},
 \cite{phtrans},\cite{ssv}, \cite{mova}, \cite{wit3} or properties
of solitonic solutions of cosmological interest, such as extreme
black holes \cite{malda} and their entropy.
A major mathematical tool in these studies is the structure of
supergravity theories in diverse dimensions \cite{sase} and with different
numbers of supersymmetries.
These theories have a central extension that gives an apparent
violation of the Haag-Lopuszanski-Shonius theorem \cite{hls}, since they include
``central charges'' that are not Lorentz-invariant \cite{tow}.
However, these charges are important because they are related to
$p$-extended objects (for charges with $p$ antisymmetrized indices)
whose dynamics is now believed to be as fundamental as that of points
and strings \cite{dual}.
In fact point-like and string-like BPS states can be obtained by
wrapping $p$ (or $p - 1$) of the dimensions of a $p$-extended object
living in $D$ dimensions when $d\geq p$ dimensions have been
compactified.
It is the aim of this paper to give a detailed analysis of central
extensions of different supergravities existing in arbitrary
dimensions $4\leq D < 10$ in a unified framework and to study the moduli dependence of the
BPS mass per unit of $p$-volume of generic BPS $p$-branes existing in
a given theory.
A basic tool in our investigation will be an exploitation of ``duality symmetries''
\cite{cfs}, \cite{fsz} (rephrased nowadays
as U--duality) of the underlying supergravity theory which, for a theory with more than 8
supercharges, takes the form of a discrete subgroup of the continuous isometries of the scalar
field sigma model of the theory \cite{ht}.
Duality symmetries, which rotate electric and magnetic charges, correspond, in a string context,
 to certain perturbative or nonperturbatives symmetries of the BPS spectrum,
 playing a crucial role in the study  of string dynamics.

The basic focus of our approach is that the central extension of the
supersymmetry algebra \cite{hls}\cite{fesaz} is encoded in the supergravity transformation
rules.
The latter can be derived from supersymmetric Bianchi identities,
even if the complete lagrangian has not yet been derived.
Of course a careful study of these identities also allows a complete
determination of the lagrangian, whenever it exists.
Among the novelties of this analysis is a new formulation of $D=4$, $N$-extended
theories with $N > 2$, in which a manifest symplectic formulation is used.
In particular all these theories have in common a flat symplectic bundle which encodes
the differential relations among the symplectic sections and therefore among the central
and matter charges.
In this respect the $N=2$ case, related to Special Geometry \cite{str}, \cite{voi}, simply differs from the
$N>2$ cases by the fact that the base space is not necessarily a coset space. This is related to
the physical fact that $N=2$ Special Geometry suffers quantum corrections.
For higher dimensional theories, relations between central and matter charges for different
$p$-extended objects can be derived in in a way strictly analogous to that presented in this lecture in
$D=4$ \cite{noi}.
\par
The application of this geometrical setting to black-hole physics, in
particular to the determination of the black-hole entropy and fixed scalars,
see \cite{uinv4} is described in the second part of these lectures.

\section{Extended supergravities and their relations with superstrings,
M-theory and F-theory}
It is worth while to recall the various compactifications of
superstrings in $4 \leq D < 10$ as well as of M-theory and their
relation to extended supergravities and their duality symmetries.
In the string context the latter symmetries are usually called S, T and U--dualities.
S--duality means exchange of small with large coupling constant, i.e. strong--weak coupling duality.
T--duality indicates the exchange of small with large volume of compactification while U--duality refers
to the exchange of NS with RR scalars.
The major virtue of space-time supersymmetry is that it links toghether these dualities; often
some of them are interchanged in comparing dual theories in the nonperturbative
regime.
In this lecture we will only consider compactifications on smooth
manifolds since the analysis is otherwise more complicated (and
richer) due to additional states concentrated at the singular points
of the moduli space.
The key ingredient to compare different theories in a given
space-time dimension is Poincar\'e duality, which converts a theory
with a $(p+ 2)$-form into one with a $(D-p-2)$-form (and inverse
coupling constant).
For example, at $D=9$ Poincar\'e duality relates 4- and 5-forms, at
$D=7$ \cite{witten} 3- and 4-forms and at $D=5$ 2- and 3-forms  \cite{schw}.
These relations are closely related to the fact that type IIA and
type IIB are T-dual at $D=9$ \cite{dhs}, heterotic on $T_3$ is dual to M-theory
on $K3$ at $D=7$ \cite{witten} and heterotic on $K3 \times S_1$ is dual to M-theory
on $CY_3$ at $D=5$ \cite{cad}, \cite{wit3}.
Let us consider dualities by first comparing theories with maximal
supersymmetry (32 supersymmetries).
An example is the duality between M-theory on $S_1$ at large radius and
type IIA in $D=10$ at strong coupling.
For the sequel we will omit the regime where these theories should be
compared.
We will just identify their low-energy effective action
including BPS states.
Further compactifying type IIA on $S_1$, it becomes equivalent to IIB
on $S_1$ with inverse radius.
This is the T-duality alluded to before.
It merely comes by Poincar\'e duality, exchanging the five form of
one theory (IIB) with the 4-form of the other theory (IIA).
The interrelation between M-theory and type IIA and type IIB theories
at $D \leq 9$ explains most of the symmetries of all maximally
extended supergravities.
At $D=8$ we have a maximal theory with U-duality \cite{ht} group $Sl(3,\ZZ)
\times Sl(2,\ZZ)$.
The  $Sl(3,\ZZ)$ has a natural interpretation from an
M-theory point of view, since the $D=8$ theory is M-theory on $T_3$.
On the other hand, the additional $Sl(2,\ZZ) $ which acts on the
4-form and its dual has a natural interpretation from the type IIB
theory on $T_2$, in which the $Sl(2,\ZZ)$ is related to the complex
structure of the 2-torus \cite{kuva}.
At $D=7$ the $ Sl(5,\ZZ) $ U-duality has no obvious interpretation unless we move
to an F--theory setting \cite{kuva}
This is also the case for $D=6,5,4$, where the U-duality groups are
$O(5,5;\ZZ)$, $E_{6,6}(\ZZ)$ and  $E_{7,7}(\ZZ)$  respectively.
However, they share the property that the related continuous group has, as maximal compact
subgroup, the automorphism group of the supersymmetry algebra, i.e.
$Sp(4)$, $Sp(4) \times Sp(4) $, $Usp(8)$ and $SU(8)$ respectively for
$D=7,6,5$ and 4.
The U--duality group for any $D$ corresponds to the series of $E_{11-D}$ Lie algebras
whose quotient with the above automorphism group of the supersymmetry algebra
provides the local description of the scalar fields moduli space
 \cite{cj}.
Recently, a novel way to unravel the structure of the U--duality groups
 in terms of solvable Lie algebras has been proposed in   \cite{solv}.
Moving to theories with lower (16) supersymmetries, we start to have
dualities among heterotic, M-theory and type II theories on manifolds
preserving 16 supersymmetries.
For $D=7$, heterotic theory on $T_3$ is ``dual'' to M-theory on $K3$
in the same sense that M-theory is ``dual'' to type IIA at $D=10$.
Here the coset space  $O(1,1) \times {O(3,19) \over O(3)\times O(19)}$
identifies the dilaton and Narain lattice of the heterotic string
with the classical moduli space of $K3$, the dilaton in one theory
being related to the volume of $K3$ of the other theory \cite{witten}.
The heterotic string on $T_4$ is dual to type IIA on $K3$.
Here the coset space $O(1,1) \times {O(4,20) \over O(4)\times O(20)}$
identifies the Narain lattice with the ``quantum'' moduli space of $K3$
(including torsion).
The $O(1,1)$ factor again relates the dilaton to the $K3$ volume.
A similar situation occurs for the theories at $D=5$.
At $D=4$ a new phenomenon occurs since the classical moduli space
${SU(1,1) \over U(1)}\times {O(6,22) \over O(6)\times O(22)}$
interchanges S-duality of heterotic string with T-duality of type IIA
theory and U-duality of type IIB theory  \cite{dlr}.
If we compare theories with 8 supersymmetries, we may at most start
with $D=6$.
On the heterotic side this would correspond to $K3$ compactification.
However at $D=6$ no M-theory or type II correspondence is possible
because we have no smooth manifolds of dimension 5 or 4 which
reduce the original supersymmetry (32) by one quarter.
The least we can do is to compare theories at $D=5$, where heterotic
theory on $K3 \times S_1$ can be compared \cite{cad} and in fact is dual, to
M-theory on a Calabi-Yau threefold which is a $K3$ fibration \cite{klema}.
Finally, at $D=4$ the heterotic string on $K3 \times T_2$ is dual to type
IIA (or IIB) on a Calabi-Yau threefold (or its mirror) \cite{cdfv}, \cite{fehastva}, \cite{kava}.
It is worth noticing that these ``dualities'' predict new BPS states as
well as they identify perturbative BPS states of one theory with
non-perturbative ones in the dual theory.
A more striking correspondence is possible if we further assume the
existence of 12 dimensional F-theory such that its compactification
on $T_2$ gives type IIB at $D=10$  \cite{vafa}.
In this case we can relate the heterotic string on $T_2$ at $D=8$ to
F-theory on $K3$ and the heterotic string at $D=6$ on $K3$ with
F-theory on a Calabi-Yau threefold  \cite{mova}, \cite{femisa}.
To make these comparisons one has to further assume that the smooth
manifolds of F-theory are elliptically fibered \cite{mova}.
An even larger correspondence arises if we also include type I strings \cite{type1}
and D-branes \cite{dbrane} in the game.
However we will not further comment on the other correspondences
relating all string theories with M and F theory.

\section{Duality symmetries and central charges in diverse dimensions}
\subsection{The general framework}
In this section we study the general group theoretical framework for
the construction of the graviphotons and matter vectors appearing in
the transformation laws of the fermion fields, the central and
matter charges and the relations among them in any dimensions.
After these preliminaries, we concentrate on the four dimensional
theories, giving the details of the construction in this case.
The details for the higher dimensional cases can be found in
\cite{noi}.

 All supergravity
theories contain scalar fields whose kinetic Lagrangian is described
by $\sigma$--models of the form $G/H$, with the exception of $D=4$, $N=1,2$ and $D=5$, $ N=2$.\\
Here $G$ is a non compact group acting as an isometry group on the
scalar manifold while $H$, the isotropy subgroup, is of the form:
\begin{equation}
H=H_{Aut} \otimes H_{matter}
\end{equation}
 $H_{Aut}$ being the automorphism group of the supersymmetry algebra
 while $H_{matter}$ is related to the matter multiplets.
 (Of course $H_{matter}=\bfone$ in all cases where supersymmetric matter
 doesn't exist, namely $N>4$ in $D=4,5$ and in general in all
 maximally extended supergravities).
The coset manifolds $G/H$ and the automorphism groups for various
 supergravity theories for any $D$ and $N$  can be found in the literature
 (see for instance
\cite{sase}, \cite{bibbia}).
 As it is well known, the group G acts linearly on the ($n=p+2$)--forms
 field strengths $H^\Lambda _{a_1\cdots a_n}$ corresponding to the
 various ($p+1$)--forms appearing in the gravitational and matter
 multiplets. Here and in the following the index $\Lambda$ runs over
 the dimensions of some representation of the duality group $G$.
The true duality symmetry (U--duality), acting on integral  quantized electric
and magnetics charges,
 is  the restriction of  the continuous group $G$ to the integers
 \cite{ht}. The moduli space of these theories is $ G(\ZZ)\backslash G/H$.
 \par
 All the properties of the given supergravity theories for fixed $D$
 and $N$ are completely fixed in terms of the geometry of $G/H$,
 namely in terms of the coset representatives $L$ satisfying the
 relation:
 \begin{equation}
 L(\phi^\prime) =g L(\phi) h (g,\phi)
\end{equation}
where $g\in G$, $h\in H$  and $\phi ^\prime =   \phi ^\prime
 (\phi)$,
 $\phi$ being the coordinates of $G/H$.
Note that the  scalar fields in $G/H$ can be assigned, in the linearized theory,  to linear representations
 $R_H$ of the local isotropy group  $H$ so that dim $R_H$ = dim $G$ $-$ dim $H$ (in the full theory, $R_H $ is the representation
 which the vielbein of $G/H$ belongs to).
\par
 As explained in the following, the kinetic metric for
 the ($p+2$)--forms $H^\Lambda$ is fixed in terms of $L$ and the
 physical field strengths of the interacting theories are "dressed"
 with scalar fields in terms of the coset representatives.
 This allows us to write down the central charges associated to the
 ($p+1$)--forms in the gravitational multiplet in a neat way in terms
 of the geometrical structure of the moduli space.
In an analogous way also the matter ($p+1$)--forms of the matter
multiplets give rise to charges which, as we will see, are closely
related to the central charges. Note that when $p>1$ the central
charges do not appear in the usual supersymmetry algebra, but in the
extended version of it containing central generators $Z_{a_1 \cdots
a_p}$ associated to $p$--dimensional extended objects ($a_1 \cdots
a_p$ are a set of space--time antisymmetric
Lorentz indices) \cite{df, vpvh, tow,
ach, bars}
\par
Our main goal is to write down the explicit form of the dressed
charges and to find relations among them analogous to those worked
out in $D=4$, $N=2$  by means of the Special Geometry relations \cite{cdf}\cite{cdfv}.
\par
To any ($p+2$)--form $H^\Lambda$ we may associate a magnetic charge (($D-p-4$)--brane)
and  an
electric ($p$--brane) charge given respectively by:
\begin{equation}
g^\Lambda = \int _{S^{p+2}} H^\Lambda
\qquad \qquad
e_\Lambda = \int _ {S^{D-p-2}}  \cG _\Lambda
\end{equation}
where $\cG_{\Lambda}= {\partial \cL \over \partial H^\Lambda}$.
\par
These charges however are not the physical charges of the interacting
theory; the latter ones can be computed by looking at the
transformation laws of the fermion fields, where the physical
field-strengths appear dressed with the scalar fields \cite{noi1},\cite{noi}.
Let us first introduce the central charges:
they are associated to the dressed ($p+2$)--forms $T^i_{AB}$ appearing
in the supersymmetry transformation law of the gravitino 1-form.
Quite generally we have, for any $D$ and $N$:
\begin{equation}
\delta \psi_A = D\epsilon_A + \sum_{i} c_iT^i _ {AB\vert a_1\cdots a_{n_i}}\Delta^{a a_1\cdots a_{n_i}}
\epsilon^B V_a+ \cdots
\label{tragra}
\end{equation}
where:
 \begin{equation}
\Delta_{a a_1\cdots a_n}=\left( \Gamma _{a a_1 \cdots
a_{n}} - {n \over n-1} (D-n-1)\delta^a_{[a_1} \Gamma_{a_2\cdots
a_{n}]} \right).
\label{delta}
\end{equation}
Here $D$ is the covariant derivative in terms of the space--time spin connection
and the composite connection of the automorphism group $H_{Aut}$,
 $c_i$ are coefficients fixed by supersymmetry, $V^a$ is the
 space--time vielbein, $A=1,\cdots,N$ is the index acted on by the
 automorphism group, $\Gamma_{a_1\cdots a_n}$ are $\gamma$--matrices
 in the appropriate dimensions, and the sum runs over all the ($p+2$)--forms
 appearing in the gravitational multiplet. Here and in the following
 the dots denote trilinear fermion terms.
Each $n$-form field-strength $T^i_{AB}$ is constructed by dressing the bare
field-strengths  $H^\Lambda$ with the coset representative $L(\phi)$ of $G/H$,
$\phi$ denoting a set of coordinates of $G/H$.
In particular, for any $p$, except for $D/2 = p+2$, we have:
 \begin{equation}
T^i_{AB} = L_{AB \Lambda_i} (\phi)
H^{\Lambda_i}
\end{equation}
where we have used the following decomposition of $L$:
\begin{equation}
L=(L^\Lambda_{AB}, L^\Lambda_I) \quad\quad L^{-1}=(L_{AB \Lambda}, L_{I \Lambda})
\label{defl}
\end{equation}
Here $L^\Lambda_\Sigma$ belongs to the representation of $G$ under which the ($p+2$)--forms $H^\Lambda$
transform irreducibly and the couple of indices $AB$ and $I$ refer to the
transformation properties of $L$  under the right action of $H_{Aut} \times H_{matter}$.
More precisely, the couple of indices $AB$ transform in the twofold tensor representation of $H_{Aut}$, which
in general is a  $Usp(N)$
group (except in  $D=8$, $N=1$ and $D=9$, $N=2$ theories where $H_{Aut}$ is $U(1)$ or $SU(2)\times U(1)$ respectively).
and $I$ is an index in the fundamental representation of $H_{matter}$ which in general is an orthogonal group.
Note that in absence of matter multiplets
$L\equiv ( L^\Lambda_{\ AB})$.
In all these cases ($D/2 \neq p+2$) the kinetic matrix of the ($p+2$)--forms $H^\Lambda$
is given in terms of the coset representatives as follows:
\begin{equation}
{1 \over 2} L_{AB\Lambda}L^{AB}_{\ \ \Sigma} - L_{I\Lambda}L^{I}_{\ \Sigma}=
\cN_{\Lambda\Sigma}\label{ndil}
\end{equation}
 with the indices of $H_{Aut}$  raised
 and lowered with the appropriate
metric of $H_{Aut}$  in the given representation.
For maximally extended supergravities
 $\, \cN_{\Lambda\Sigma} = L_{AB\Lambda}L^{AB}_{\ \ \Sigma}$.
Note that both for matter coupled and maximally extended
supergravities we have:
\begin{equation}
L_{\Lambda AB} = \cN_{\Lambda\Sigma}L^{\Sigma}_{\ AB}
\label{invl}
\end{equation}
When $G$ contains an orthogonal factor $O(m,n)$, what happens
 for matter coupled supergravities in $D=5,7,8,9$, where $G=O(10-D,n)  \times O(1,1)$
 and in all the matter coupled $D=6$ theories,
 the coset
representatives of the orthogonal group satisfy:
\begin{eqnarray}
L^t\eta L = \eta &\to &  L_{r\Lambda}L_{r \Sigma} -  L_{I\Lambda}L_{I \Sigma}=
\eta_{\Lambda\Sigma} \label{etadil}\\
L^t L = \cN &\to & L_{r \Lambda}L_{r \Sigma} +  L_{I\Lambda}L_{I \Sigma}=
\cN_{\Lambda\Sigma}
\end{eqnarray}
 where $\eta=\pmatrix{\bfone_{m\times m} & 0 \cr 0 & -\bfone_{n\times n}\cr}$
 is the $O(m,n)$ invariant metric  and $A=1,\cdots,m$; $I = 1,\cdots,n$
(In particular,  setting the matter to zero, we have in these cases
$\cN_{\Lambda\Sigma}= \eta_{\Lambda\Sigma}$).\\
In these cases we have:
\begin{equation}
 L^\Lambda_{\ AB}= L^\Lambda_{\ r} (\gamma^r)_{AB},
\end{equation}
$(\gamma^r)_{AB}$ being the $\gamma$-matrices intertwining between orthogonal and
$USp(N)$ indices.
\par
When $D$ is even and $D/2 = p+2$ the previous formulae in general require modifications,
since in that case we have the
 complication  that the action of $G$ on the $p+2 =D/2$-forms ($D$ even) is
 realized through the embedding of $G$ in $Sp(2n,\IR)$ ($p$ even) or $O(n,n)$
($p$ odd)
 groups \cite{gz}, \cite{cfg}.\\
This happens for $D=4$, $N>1$, $D=6$ , $N=(2,2)$ and the maximally extended $D=6$ and $D=8$ supergravities.
\footnote{The 6 dimensional theories $N=(2,0)$ and $N=(4,0)$ do not require such embedding
since the 3--forms $H^\Lambda$ have definite self--duality and no lagrangian exists.
For these theories the formulae of the previous odd dimensional cases are valid.}
The necessary modifications for the embedding are worked out in section $3.2$.
\par
Coming  back to the case $D/2 \neq p+2$, it is now straightforward to compute  the central charges.
\par
Indeed, the magnetic central charges for BPS saturated ($D-p-4$)--branes
can be now defined
(modulo numerical factors to be fixed in each theory) by integration
of the dressed
field strengths as follows:
\begin{equation}
Z^{(i)}_{(m) AB} = \int _{S^{p+2}}T^i_{ AB}=\int _{S^{p+2}}L_{\Lambda_i AB}(\phi) H^{\Lambda_i}=
 L_{\Lambda_i AB}(\phi_0) g^{\Lambda_i}
 \label{carma}
\end{equation}
where $\phi_0$ denote the $v.e.v.$ of the scalar fields, namely
$\phi_0 = \phi(\infty)$ in a given background.
The corresponding electric central charges are:
 \begin{equation}
Z^{(i)}_{(e) AB} = \int _{S^{D-p-2}}L_{\Lambda_i AB}(\phi)
^{\ \star}H^{\Lambda_i}= \int _{S^{D-p-2}} \cN _{\Lambda_i\Sigma_i}
L^{\Lambda_i}_{\  AB}(\phi)
^{\ \star}H^{\Sigma_i}=
 L^{\Lambda_i}_{\ AB}(\phi_0) e_{\Lambda_i}
\end{equation}
 These formulae make it explicit that $L^\Lambda_{\ AB}$ and
 $L_{\Lambda AB}$ are related by electric--magnetic duality via the
 kinetic matrix.
 \par
 Note that the same field strengths $T^i_{AB}$ which appear in the gravitino
 transformation laws are also present in the dilatino transformation laws
  in the following way:
  \begin{equation}
\delta \chi_{ABC} = \cdots +
\sum_{i} b_i L_{\Lambda_i AB} (\phi)
H^{\Lambda_i} _ {a_1\cdots a_{n_i}}\Gamma^{ a_1\cdots a_{n_i}} \epsilon_C+\cdots
\label{tradil}
\end{equation}
\vskip5mm
\par
 In an analogous way, when vector multiplets are present,
 the matter vector field
 strengths are dressed with the columns $L_{\Lambda I}$
 of the coset element (\ref{defl})
 and they
 appear in the transformation laws of the gaugino fields:
  \begin{equation}
\delta \lambda^I_{A} = c_1 \Gamma^a P^I_{AB , i}
\partial_a \phi^i \epsilon^B +
 c_2 L_{\Lambda}^{\ I} (\phi)
F^{\Lambda} _ {ab}\Gamma^{ ab} \epsilon_A  + \cdots
\label{tragau}
\end{equation}
  where  $ P^I_{AB }=P^I_{AB,i }d\phi^i$ (see eq. (\ref{dllp}) in the following)
 is the vielbein of the coset manifold
  spanned by the scalar fields of the vector multiplets,
 $F^\Lambda_{ab}$ is the field--strength of the matter photons
 and $c_1, c_2$  are  constants fixed by supersymmetry (in $D=6$, $N=(2,0)$ and $N=(4,0)$ the 2--form
 $F^{\Lambda}_{ab} \Gamma^{ab}$ is replaced by the 3--form
  $H^{\Lambda}_{abc} \Gamma^{abc}$).
In the same way as for central charges, one finds the
magnetic matter charges:
\begin{equation}
  Z_{(m) A} ^{\ I} = \int _{S^{p+2}} L_\Lambda^{\ I} F^\Lambda
   =   L_\Lambda^{\ I} (\phi_0) g^\Lambda
\end{equation}
while the electric matter charges are:
\begin{equation}
Z_{(e) I} = \int _{S^{D-p-2}}L_{\Lambda I}(\phi)
^{\ \star}F^{\Lambda}= \int _{S^{D-p-2}} \cN _{\Lambda\Sigma}
L^{\Lambda}_{\  I}(\phi)
^{\ \star}F^\Sigma =
 L^{\Lambda}_{\ I}(\phi_0) e_{\Lambda}
\end{equation}
 \par
 The important fact to note is that the central charges and matter
 charges satisfy relations and sum rules analogous to those derived
 in $D=4$, $N=2$ using Special Geometry techniques which we review
 in the following \cite{cdf}.
They are inherited from the
 properties of the coset manifolds $G/H$, namely from the
 differential and algebraic properties satisfied by the coset
 representatives $L^\Lambda_{\ \Sigma}$.
Indeed, for a general coset manifold we may introduce the
left-invariant 1-form $\Omega=L^{-1} d L$ satisfying the
relation (see for instance \cite{bibbia}):
\begin{equation}
d \Omega + \Omega \wedge \Omega =0
\label{mc}
\end{equation}
 where \begin{equation}
\Omega=\omega^i T_i + P^\alpha T_\alpha
\label{defomega}
\end{equation}
  $T_i, T_\alpha$ being the generators of $G$ belonging respectively to the Lie
  subalgebra $\IH$ and to the coset space algebra $\IK$ in the Cartan
  decomposition
  \begin{equation}
\IG = \IH + \IK
\label{hk}
\end{equation}
 $\IG$ being the Lie algebra of $G$. Here $\omega^i$ is the $\IH$
 connection and $P^\alpha$, in the representation $R_H$ of $H$,  is the vielbein of $G/H$.
Since in all the cases we will consider $G/H$ is a symmetric space
($[\IK , \IK ] \subset \IH$), $\omega^i C_i^{\ \alpha\beta}$
($C_i^{\ \alpha\beta}$ being the structure constants of $G$)
can be identified
with the Riemannian spin connection of $G/H$.
\par
Suppose now we have a  matter coupled theory. Then, using
the decomposition (\ref{hk}), from
(\ref{mc}) and (\ref{defomega}) we get:
\begin{equation}
dL^\Lambda_{\ AB} = {1 \over 2}L^\Lambda_{\ CD} \omega^{CD}_{\ \ AB} +
L^\Lambda_{\ I} P^I_{AB}
\label{cosetmc}
\end{equation}
where $P^I_{AB}$ is the vielbein on $G/H$ and
 $\omega^{CD}_{\ \ AB}$ is the $\IH$--connection in the given
 representation.
It follows:
\begin{equation}
\nabla^{(H)} L^\Lambda_{\ AB}= L^\Lambda_{\ I} P^I_{AB}
\label{dllp}
\end{equation}
where the derivative is covariant with  respect to the
$\IH$--connection  $\omega^{CD}_{\ \ AB}$.
Using the definition of the magnetic dressed charges given in
(\ref{carma}) we obtain:
\begin{equation}
\nabla^{(H)} Z_{AB}= Z_{I} P^I_{AB}
\label{dz}
\end{equation}
 This is a prototype of the formulae one can derive in the various
 cases for matter coupled supergravities \cite{noi}.
 To illustrate one possible application of this kind of formulae let
 us suppose that in a given background preserving some number of
 supersymmetries $Z_I=0$ as a consequence of $\delta\lambda^I_A=0$.
 Then we find:
 \begin{equation}
\nabla^{(H)} Z_{AB}=0 \to d(Z_{AB} \bar Z^{AB} )=0
\end{equation}
 that is the square of the
 central charge reaches an extremum with respect to the
 $v.e.v.$ of the moduli fields. Backgrounds
with such fixed scalars describe the horizon geometry of extremal black holes and behave
as attractor points for the scalar fields evolution in the black hole geometry
 \cite{fks},\cite{feka1},\cite{feka2}.
\par
 For the maximally extended supergravities there are no matter
 field--strengths and the previous differential relations become
differential relations among central charges only.
As an example, let us consider $D=5$, $N=8$ theory. In this case the Maurer--Cartan equations become:
\begin{equation}
dL^{\Lambda}_{\ AB} ={1 \over 2} L^{\Lambda}_{\ CD} \Omega^{CD}_{\ \ AB}+
{1 \over 2}\bar L^{\Lambda CD}  P_{CD AB}
\end{equation}
 where the coset representative is taken in the
$27 \times 27$ fundamental  representation of $E_6$, $\Omega^{CD}_{\ \ AB}=
 2 Q^{[A}_{\ [c}\delta^{B]}_{D]}$, $AB$ is  a couple of antisymmetric symplectic--traceless
$USp(8)$ indices, $Q^A_{\ B}$ is the $USp(8)$ connection and the vielbein $P_{CDAB}$
is antisymmetric, $\IC_{AB}$--traceless and pseudo--real.
Note that $(L^\Lambda_{\ CD})^*= L^ {\Lambda CD}$.
 Therefore we get:
  \begin{equation}
\nabla^{(H)} L^\Lambda_{\ AB}={1 \over 2} \bar L^{\Lambda CD} P_{CDAB}
\label{dllp2}
\end{equation}
   that is:
  \begin{equation}
\nabla^{(H)} Z_{AB}= {1 \over 2}\bar Z^{CD} P_{CD AB}
\label{dz2}
\end{equation}
 This relation implies that the vanishing of a subset of central
 charges forces the vanishing of the covariant derivatives of some
 other subset.
 Typically, this happens in some
 supersymmetry preserving backgrounds where
 the requirement $\delta\chi_{ABC}=0$ corresponds to the vanishing of
 just a subset of central charges.
 Finally, from the coset representatives relations
 (\ref{ndil}) (\ref{etadil}) it
 is immediate to obtain sum rules for the central and matter charges
 which are
the counterpart of those found in $N=2$, $D=4$ case using Special
Geometry \cite{cdf}.
 Indeed, let us suppose e.g. that the group $G$ is
 $G=O(10-D,n)\times O(1,1)$, as it
 happens in general for all the minimally extended supergravities in
 $7 \leq D \leq 9$,  $D=6$ type $IIA$ and
 $D=5$, $N=2$.
  The coset representative is now a tensor product $L \to e^\sigma L$, where
  $e^\sigma$ parametrizes the $O(1,1)$ factor.\\
  We have, from   (\ref{etadil})
 \begin{equation}
L^t \eta L =\eta
\end{equation}
 where $\eta$ is the invariant metric of $O(10-D,n)$ and  from  (\ref{ndil})
 \begin{equation}
e^{-2\sigma}(L^{t} L)_{\Lambda\Sigma} =\cN_{\Lambda\Sigma}.
\end{equation}
 Using the decomposition (\ref{hk}) one finds:
 \begin{equation}
{1 \over 2} Z_{AB} Z_{AB} - Z_I Z_I = g^\Lambda \eta _{\Lambda\Sigma} g^\Sigma
e^{-2\sigma}
\end{equation}
   \begin{equation}
{1 \over 2} Z_{AB} Z_{AB} + Z_I Z_I = g^\Lambda \cN _{\Lambda\Sigma} g^\Sigma
\end{equation}
 In more general cases analogous relations of the same kind can be
 derived \cite{noi}.

 \subsection{The embedding procedure in $D=4$}
In this subsection we work out the modifications to the formalism
developed in the previous subsection
in the case $D/2 = p+2$, which derive from the embedding procedure
\cite{gz} of the group $G$
in $Sp(2n,\IR)$ ($p$ even) or in $O(n,n)$ ($p$ odd).
We concentrate on $D=4$, while for $D=6,8$ we
refer to ref. \cite{noi}.
Furthermore we show that the flat symplectic bundle formalism of the $D=4$,
$N=2$ Special Geometry case \cite{str}, \cite{voi}
can be extended to $N>2$ theories.
The $N=2$ case differs from the other higher $N$ extensions by the fact that the base space
of the flat symplectic bundle is not in general a coset manifold.
\par
Let us analyze the structure of the four dimensional theories.
\par
 In $D=4$, $N>2$ we may decompose the vector field-strengths in self-dual and
 anti self-dual parts:
 \begin{equation}
F^{\mp} = {1\over 2}(F\mp {\rm i} ^{\ \star} F)
\end{equation}
  According to the Gaillard-Zumino construction, $G$ acts on the
  vector $(F^{- \Lambda},\cG^{-}_\Lambda)$
  (or its complex conjugate) as a subgroup of
  $Sp(2 n_v,\IR)$ ($n_v$ is the number of vector fields)
with duality transformations interchanging electric and magnetic
 field--strengths:
 \begin{equation}
{\cal S}
\left(\matrix {F^{-\Lambda} \cr
\cG^-_\Lambda\cr}\right)=
\left(\matrix {F^{-\Lambda} \cr
\cG^-_\Lambda\cr}\right)^\prime
\end{equation}
 where:
 \begin{eqnarray}
\cG^-_\Lambda&=&\bar \cN_{\Lambda\Sigma}F^{-\Sigma}\nonumber\\
 \cG^+_\Lambda&=& \cN_{\Lambda\Sigma}F^{+\Sigma}
 \label{defg}
\end{eqnarray}
\begin{equation}
 {\cal S}=\left( \matrix{A& B\cr C & D \cr}\right)\in G \subset Sp(2 n_v,\IR) \to \quad \left\{\matrix{
A^t C -C^t A &=&0 \cr B^t D -D^t B &=& 0 \cr A^t D -C^t B &=&1 } \right.
\label{defs}
\end{equation}
  and $\cN_{\Lambda\Sigma}$, is the symmetric matrix appearing in the kinetic
  part of the vector Lagrangian:
  \begin{equation}
  \cL_{kin}= {\rm i}\bar \cN_{\Lambda\Sigma}F^{-\Lambda} F^{-\Sigma} + h.
  c.
  \end{equation}
If $L(\phi)$ is the coset representative of $G$ in some representation, $S$ represents the embedded coset representative
belonging to $Sp(2n_v,\IR)$ and in each theory, $A,B,C,D$ can be constructed in terms of $L(\phi)$.
Using a complex basis in the vector space of $Sp(2 n_v)$, we may
rewrite the
symplectic matrix as an $Usp(n_v,n_v)$ element:
\begin{equation}
U = {1 \over \sqrt{2}}\pmatrix{f+{\rm i}h & \bar f+{\rm i}\bar h \cr
f-{\rm i}h &\bar f-{\rm i}\bar h \cr} =
  \cA^{-1} S \cA
  \label{defu}
\end{equation}
where:
\begin{eqnarray}
  f&=&{1 \over \sqrt{2}} (A-{\rm i} B) \nonumber\\
h &=&{1 \over \sqrt{2}} (C-{\rm i} D) \nonumber\\
 \cA &=& \pmatrix{1 & 1 \cr -{\rm i} & {\rm i}\cr}
\end{eqnarray}
The requirement $ {U} \in Usp(n_v, n_v)$ implies:
 \begin{equation}
\left\lbrace\matrix{{\rm i}(f^\dagger h - h^\dagger f) &=& \bfone \cr
(f^\dagger \bar h - h^\dagger \bar f) &=& 0\cr} \right.
\label{specdef}
\end{equation}
The $n_v\times n_v$ subblocks of U are submatrices $f,h$ which can be decomposed with respect to the
isotropy group $H_{Aut} \times H_{matter}$ in the same way as $L$ in equation (\ref{defl}), namely:
\begin{eqnarray}
  f&=& (f^\Lambda_{AB} , f^\Lambda _I) \nonumber\\
h&=& (h_{\Lambda AB} , h_{\Lambda I})
\label{deffh}
\end{eqnarray}
where $AB$ are indices in the antisymmetric representation of $H_{Aut}= SU(N) \times U(1)$ and
$I$ is an index of the fundamental representation of $H_{matter}$.
Upper $SU(N)$ indices label objects in the complex conjugate representation of $SU(N)$:
$(f^\Lambda_{AB})^* = f^{\Lambda AB}$ etc.
\par
Note that we can consider $(f^\Lambda_{AB}, h_{\Lambda AB})$ and $(f^\Lambda_{I}, h_{\Lambda I})$
as symplectic sections of a $Sp(2n_v, \IR)$ bundle over $G/H$.
We will see in the following that this bundle is actually flat.
The real embedding given by $S$ is appropriate for duality transformations  of $F^\pm$
 and their duals $\cG^\pm$, according to equations (\ref{defs}), (\ref{defg}), while
the complex embedding in the matrix $U$ is appropriate in writing down the fermion transformation
laws and supercovariant field--strengths.
The kinetic matrix $\cN$, according to Gaillard--Zumino \cite{gz}, turns out to be:
\begin{equation}
\cN= hf^{-1}, \quad\quad \cN = \cN^t
\label{nfh-1}
\end{equation}
 and transforms projectively under $Sp(2n_v ,\IR)$ duality rotations:
 \begin{equation}
\cN^\prime = (C+ D \cN) (A+B\cN)^{-1}
\end{equation}
  By using (\ref{specdef})and (\ref{nfh-1}) we find that
   \begin{equation}
   (f^t)^{-1} = {\rm i} (\cN - \bar \cN)\bar f
\end{equation}
which is the analogous of equation (\ref{invl}), that is
 \begin{eqnarray}
  f_{AB \Lambda} &\equiv& (f^{-1})_{AB \Lambda} =
{\rm i} (\cN - \bar \cN)_{\Lambda\Sigma}\bar f^\Sigma_{AB}\\
f_{I \Lambda} &\equiv& (f^{-1})_{I \Lambda} =
{\rm i} (\cN - \bar \cN)_{\Lambda\Sigma}\bar f^\Sigma_{I}
\end{eqnarray}
It follows that the dressing factor $(L^\Lambda)^{-1}=(L_{\Lambda AB}, L_{\Lambda I})$
in equation (\ref{tragra}) which was given by the inverse coset representative
in the defining representation of $G$ has to be replaced by the analogous
inverse representative $(f_{\Lambda AB}, f_{\Lambda I})$ when, as in the
present $D=4$ case, we have to embed $G$ in $Sp(2n,\IR)$.
 As a consequence, in the transformation law of gravitino (\ref{tragra}), dilatino (\ref{tradil})
 and gaugino (\ref{tragau})
 we perform the following replacement:
 \begin{equation}
(L_{\Lambda AB}, L_{\Lambda I}) \to  (f_{\Lambda AB}, f_{\Lambda I})
\end{equation}
In particular, the dressed graviphotons and matter self--dual field--strengths take the
symplectic invariant form:
\begin{eqnarray}
T^-_{AB}&=& -{\rm i} (\bar f^{-1})_{AB \Lambda}F^{- \Lambda} = f^\Lambda_{AB}(\cN - \bar \cN)_{\Lambda\Sigma}F^{-\Sigma}=
 h_{\Lambda AB} F^{-\Lambda} -f^\Lambda_{AB} \cG^-_\Lambda  \nonumber\\
  T^-_{I}&=&-{\rm i} (\bar f^{-1})_{I\Lambda}F^{- \Lambda} =  f^\Lambda_{I}(\cN - \bar \cN)_{\Lambda\Sigma}F^{-\Sigma}=
 h_{\Lambda I} F^{-\Lambda} - f^\Lambda_{I} \cG^-_\Lambda \nonumber\\
\bar T^+_{AB} &=& (T^-_{AB})^* \nonumber\\
\bar T^+_{I} &=& (T^-_{I})^* \label{gravi}
\end{eqnarray}
(Obviously, for $N>4$, $L_{\Lambda I} = f _{\Lambda I}= T_I =0$).
 To construct the dressed charges one integrates
 $T_{AB} = T^+_{AB} + T^ -_{AB}  $ and  (for $N=3, 4$)
 $T_I = T^+_I + T^ -_I $ on a large 2-sphere.
 For this purpose we note that
 \begin{eqnarray}
 T^+_{AB} & = &  h_{\Lambda
 AB}F^{+\Lambda} - f^\Lambda_{AB} \cG_\Lambda^+  =0  \label{tiden0}\\
   T^+_I & = &  h_{\Lambda
 I}F^{+\Lambda}-f^\Lambda_{I} \cG_\Lambda^+   =0 \label{tiden}
\end{eqnarray}
as a consequence of eqs. (\ref{nfh-1}), (\ref{defg}).
Therefore we have:
\begin{eqnarray}
Z_{AB} & = & \int_{S^2} T_{AB} = \int_{S^2} (T^+_{AB} + T^ -_{AB}) = \int_{S^2} T^ -_{AB} =
  h_{\Lambda AB} g^\Lambda- f^\Lambda_{AB} e_\Lambda
\label{zab}\\
Z_I & = & \int_{S^2} T_I = \int_{S^2} (T^+_I + T^ -_I) = \int_{S^2} T^ -_I =
 h_{\Lambda I} g^\Lambda - f^\Lambda_I e_\Lambda   \quad (N\leq 4)
\label{zi}
\end{eqnarray}
where:
\begin{equation}
e_\Lambda = \int_{S^2} \cG_\Lambda , \quad
g^\Lambda = \int_{S^2} F^\Lambda \label{charges}
\end{equation}
and the sections $(f^\Lambda, h_\Lambda)$ on the right hand side now depend on the {\it v.e.v.}'s
of the scalar fields $\phi^i$.
We see that the  central and matter charges are given in this case by  symplectic invariants
and that the presence of dyons in $D=4$ is related to the
 symplectic embedding.
In the case $D/2\neq p+2$, $D$ even, or $D$ odd, we were able to derive the differential relations
(\ref{dz}), (\ref{dz2}) among
the central and matter charges using the Maurer--Cartan equations (\ref{dllp}), (\ref{dllp2}).
The same can be done in the present case using the embedded coset representative $U$.
Indeed, let $\Gamma = U^{-1} dU$ be the $Usp(n_v,n_v)$ Lie algebra left invariant one form
satisfying:
\begin{equation}
  d\Gamma +\Gamma \wedge \Gamma = 0
\label{int}
\end{equation}
In terms of $(f,h)$ $\Gamma$ has the following form:
\begin{equation}
  \label{defgamma}
  \Gamma \equiv U^{-1} dU =
\pmatrix{{\rm i} (f^\dagger dh - h^\dagger df) & {\rm i} (f^\dagger d\bar h - h^\dagger d\bar f) \cr
-{\rm i} (f^t dh - h^t df) & -{\rm i}(f^t d\bar h - h^t d\bar f) \cr}
\equiv
\pmatrix{\Omega^{(H)}& \bar \cP \cr
\cP & \bar \Omega^{(H)} \cr }
\end{equation}
where the $n_v \times n_v$ subblocks $ \Omega^{(H)}$ and  $\cP$
embed the $H$ connection and the vielbein of $G/H$ respectively .
This identification folllows from the Cartan decomposition of the $Usp(n_v,n_v)$ Lie algebra.
Explicitly, if we define the $H_{Aut} \times H_{matter}$--covariant derivative of a vector $V= (V_{AB},V_I)$ as:
\begin{equation}
\nabla V = dV - V\omega , \quad \omega = \pmatrix{\omega^{AB}_{\ \
CD} & 0 \cr 0 & \omega^I_{\ J} \cr }
\label{nablav}
\end{equation}
we have:
\begin{equation}
\Omega^{(H)} = {\rm i} [ f^\dagger (\nabla h + h \omega) - h^\dagger
(\nabla f + f \omega)] = \omega
\end{equation}
where we have used:
\begin{equation}
\nabla h = \bar \cN \nabla f ; \quad h = \cN f
\end{equation}
and  the fundamental identity (\ref{specdef}).
Furthermore, using the same relations, the embedded vielbein  $\cP$ can be written as follows:
\begin{equation}
\cP = - {\rm i} ( f^t \nabla h - h^t \nabla f) = {\rm i} f^t (\cN -
\bar \cN) \nabla f
\end{equation}
From  (\ref{defu}) and (\ref{defgamma}),  we obtain the $(n_v \times n_v)$ matrix equation:
\begin{eqnarray}
  \nabla(\omega) (f+{\rm i} h) &=& (\bar f + {\rm i} \bar h) \cP \nonumber \\
\nabla(\omega) (f-{\rm i} h) &=& (\bar f - {\rm i} \bar h) \cP
\end{eqnarray}
together with their complex conjugates. Using further the definition (\ref{deffh}) we have:
\begin{eqnarray}
  \nabla(\omega) f^\Lambda_{AB} &=&  \bar f^\Lambda_{I}  P^I_{AB} + {1 \over 2}\bar f^{\Lambda CD} P_{ABCD} \nonumber \\
\nabla(\omega) f^\Lambda_{I} &=&  {1 \over 2} \bar f^{\Lambda AB}  P_{AB I} +\bar f^\Lambda_{J}  P^J_{I}
 \label{df}
\end{eqnarray}
where we have decomposed the embedded vielbein $\cP$ as follows:
\begin{equation}
  \label{defp}
  \cP = \pmatrix{P_{ABCD} & P_{AB J} \cr P_{I AB} & P_{IJ}\cr }
\end{equation}
 the subblocks being related to the vielbein of $G/H$, $P = L^{-1} \nabla^{(H)} L$,
 written in terms of the indices of $H_{Aut} \times H_{matter}$.
Note that, since $f$ belongs to the unitary matrix $U$, we have:
$(\bar f^\Lambda _{AB},\bar f^\Lambda_I) = (f^{\Lambda AB}, f^{\Lambda I})$.
Obviously, the same differential relations that we wrote for $f$ hold true for the dual matrix $h$ as well.
\par
Using the definition of the charges (\ref{zab}), (\ref{zi}) we  then get the following differential
relations among charges:
\begin{eqnarray}
  \nabla(\omega) Z_{AB} &=&  \bar Z_{I}  P^I_{AB} +  {1 \over 2} \bar Z^{CD} P_{ABCD} \nonumber \\
\nabla(\omega) Z_{I} &=& {1 \over 2}  \bar Z^{AB}  P_{AB I} + \bar Z_{J}  P^J_{I}
 \label{dz1}
\end{eqnarray}
Depending on the coset manifold, some of the subblocks of (\ref{defp}) can be actually zero.
For example in $N=3$ the vielbein of $G/H = {SU(3,n) \over SU(3)
 \times SU(n)\times U(1)}$ \cite{maina} is $P_{I AB}$ ($AB $ antisymmetric),
 $I=1,\cdots,n;
A,B=1,2,3$ and it turns out
that $P_{ABCD}= P_{IJ} = 0$.
\par
In $N=4$, $G/H= {SU(1,1) \over U(1)} \times { O(6,n)\over O(6) \times O(n)}$ \cite{bekose}, and we have
$P_{ABCD}= \epsilon_{ABCD} P$, $P_{IJ} = \bar P \delta_{IJ}$, where $P$ is the K\"ahlerian vielbein of ${SU(1,1)\over
U(1)}$,  ($A,\cdots , D $ $SU(4)$ indices and $I,J$ $O(n)$ indices)
and $P_{I AB}$ is the vielbein of ${ O(6,n)\over O(6) \times O(n)}$.
\par
For $N>4$ (no matter indices) we have that $\cP $ coincides with the vielbein $P_{ABCD}$ of the relevant $G/H$.
\par
For the purpose of comparison of the previous formalism with the $N=2$ Special Geometry case, it is interesting
to note that, if the connection $\Omega^{(H)}$ and the vielbein $\cP$  are regarded as data of $G/H$,
then the Maurer--Cartan equations (\ref{df}) can be interpreted as an integrable system of
differential equations for a section $V=(V_{AB}, V_I, \bar V^{AB} , \bar V^I)$ of the symplectic fiber bundle
constructed over $G/H$.
Namely the integrable system:
\begin{equation}
 \nabla \pmatrix{V_{AB} \cr V_I \cr \bar V^{AB} \cr \bar V^I } = \pmatrix{0 & 0 & {1 \over 2} P_{ABCD} & P_{AB J} \cr
0 & 0 & {1 \over 2} P_{I CD} & P_{IJ} \cr
{1 \over 2}P^{ABCD} & P^{ AB J} & 0 & 0 \cr {1 \over 2}P^{I CD} & P^{IJ} & 0 & 0 \cr}
\pmatrix{V_{CD} \cr V_J \cr \bar V^{CD} \cr \bar V^J }
 \label{intsys}
\end{equation}
 has $2n$ solutions given by $V = (f^\Lambda_{\ AB},f^\Lambda_{\ I}),
 ( h_{\Lambda AB}, h_{\Lambda_I})$, $\Lambda = 1,\cdots,n$.
The integrability condition (\ref{int}) means that $\Gamma$ is a flat
connection of the symplectic bundle.
In terms of the geometry of $G/H$
this in turn implies that the $\IH$--curvature, and hence the Riemannian
curvature, is constant, being proportional to
the wedge product of two vielbein.
\par
Besides the differential relations (\ref{dz1}), the charges also satisfy sum rules quite analogous
to those found in \cite{cdf} for the $N=2$ Special Geometry case.
\par
The sum rule has the following form:
\begin{equation}
 {1 \over 2} Z_{AB} \bar Z^{AB} + Z_I \bar Z^I = -{1\over 2} P^t \cM (\cN) P
\label{sumrule}
\end{equation}
where $\cM(\cN)$ and $P$ are:
\begin{equation}
\cM = \left( \matrix{ \bfone & 0 \cr - Re \cN &\bfone\cr}\right)
\left( \matrix{ Im \cN & 0 \cr 0 &Im \cN^{-1}\cr}\right)
\left( \matrix{ \bfone & - Re \cN \cr 0 & \bfone \cr}\right)
\label{m+}
\end{equation}
\begin{equation}
P=\left(\matrix{g^\Lambda \cr e_ \Lambda \cr} \right)
\label{eg}
\end{equation}
In order to obtain this result we just need to observe that from the
 fundamental identities (\ref{specdef1}) and  from the definition of the
 kinetic matrix given in (\ref{nhf}) it  follows:
 \begin{eqnarray}
ff^\dagger &= &{\rm i} \left( \cN - \bar \cN \right)^{-1} \\
hh^\dagger &= &{\rm i} \left(\bar \cN^{-1} - \cN^{-1} \right)^{-1}\equiv
{\rm i} \cN \left( \cN - \bar \cN \right) ^{-1}\bar \cN \\
hf^\dagger &= & \cN ff^\dagger  \\
fh^\dagger & = & ff^\dagger \bar \cN
\end{eqnarray}
We note  that the matrix $\cM$ is a symplectic tensor and in this
sense it is quite analogous to the matrix $\cN _{\Lambda\Sigma} $ of
the odd dimensional cases defined in equation (\ref{ndil}).
Indeed, one sees that the matrix $\cM$ can be written as
the symplectic analogous of (\ref{ndil}):
\begin{equation}
\cM(\cN) = \pmatrix{0&\bfone \cr -\bfone & 0} \pmatrix{f \cr h \cr}
\pmatrix{f & h \cr}^\dagger
 \pmatrix{0&\bfone \cr -\bfone & 0}
\end{equation}
where  $\pmatrix{0&\bfone \cr -\bfone & 0} \pmatrix{f \cr h \cr}$ is the embedded object corresponding
to the $L$ of equation (\ref{ndil})
\vskip 5mm
The formalism we have developed so far for the $D=4$, $N>2$ theories
is completely determined by the embedding of the coset representative of $G/H$ in $Sp(2n,\IR)$
and by the $Usp(n,n)$ embedded Maurer--Cartan equations (\ref{df}).
We want now to show that this formalism, and in particular the identities (\ref{specdef}), the differential relations
among charges (\ref{dz1}) and the sum rules (\ref{sumrule}),
are completely analogous to the Special Geometry relations of $N=2$ matter coupled
supergravity \cite{spegeo}, \cite{str}.
This follows essentially from the fact that, though that the scalar manifold $\cM_{N=2}$ of the $N=2$ theory is not in
general a coset manifold, nevertheless it has a symplectic structure identical to the
$N>2$ theories.
Furthermore we will show that the analogous of the Maurer--Cartan equations of $N>2$ theories
are given in the $N=2$ case by the Picard--Fuchs equations \cite{voi} for the symplectic sections
which enter in the definition of the Special Geometry flat symplectic bundle.
\par
Indeed, let us recall that Special Geometry can be defined in terms of the holomorphic flat
vector bundle of rank $2n$ with structure group $Sp(2n,\IR)$ over a K\"ahler--Hodge manifold \cite{str}.
\par
If we introduce the Special Geometry symplectic and covariantly holomorphic section:
\begin{equation}
V \equiv \left( f^\Lambda, h_\Lambda \right) =
\left( f^\Lambda(z^i,z^{\bar \imath}), h_\Lambda (z^i,z^{\bar \imath})\right) ; \,
\, \nabla_{\bar \imath} V=0 \quad \Lambda =1,...,n
\label {defsez}
\end{equation}
and  its covariant derivative with respect to the K\"ahler
connection:
\begin{equation}
  \nabla_i V=
\left( \partial _i + {1 \over 2} \partial _i K \right)V \equiv
\left( f^\Lambda_i, h_{\Lambda i}\right) \quad i=1,\cdots , n-1
\label{dv}
\end{equation}
then, defining the $n \times n$
matrices:
\begin{equation}
f^\Lambda_\Sigma \equiv \left( f^\Lambda,  f^\Lambda_{\bar \imath} \right);
\quad
h_{\Lambda \Sigma} \equiv \left( h_ \Lambda,
 h_{\Lambda \bar \imath} \right)
\label{matrici}
\end{equation}
the set of algebraic relations of Special Geometry can  be written  in matrix form as:
\begin{equation}
\left\lbrace\matrix{{\rm i}(f^\dagger h - h^\dagger f) &=& \bfone \cr
(f^\dagger \bar h - h^\dagger \bar f) &=& 0\cr} \right.
\label{specdef1}
\end{equation}
Recalling equations (\ref{specdef}) we see that the previous relations imply that the matrix
 $U$:
\begin{equation}
U={1\over{\sqrt{2}}}\left ( \matrix{f+{\rm i} h & \bar f +
{\rm i} \bar h \cr
 f-{\rm i} h & \bar f - {\rm i} \bar h \cr} \right)
\end{equation}
belongs to $ Usp (n,n)$.
In fact if we set $f^\Lambda \to f^\Lambda \epsilon_{AB}\equiv f^\Lambda_{AB}$
and flatten the indices of $(\bar f_i, f_{\bar\imath})$ (or $(\bar h_i, h_{\bar\imath})$)
with the K\"ahlerian vielbein $P^I_i, \bar P^{\bar I}_{\bar \imath}$:
\begin{equation}
  (\bar f^\Lambda_I, f^\Lambda_{\bar I}) = (\bar f^\Lambda_iP^i_I,
  f^\Lambda_{\bar \imath}\bar P^{\bar \imath}_{\bar I} ),
\quad  \quad P^I_{i} \bar P^{\bar J}_{\bar\jmath}\eta_{I\bar J} = g_{i\bar\jmath}
\end{equation}
where $\eta_{I\bar J}$ is the flat K\"ahlerian metric and $P^i_{\ I} =
(P^{-1})^I_{\ i}$,
the relations (\ref{specdef1}) are just a particular case of equations (\ref{specdef}) since,
for $N=2$, $H_{Aut}= SU(2) \times U(1)$, so that $f^\Lambda_{AB}$ is actually an $SU(2)$ singlet.
\par
Let us now consider the analogous of the embedded Maurer-Cartan equations  of $G/H$.
Defining as before  the matrix one-form $\Gamma= U^{-1} dU$ valued in the $Usp(n,n)$ Lie algebra,
we see that the relation $d\Gamma+ \Gamma\wedge\Gamma =0$ again implies
a flat connection for the symplectic bundle over the K\"ahler--Hodge manifold.
However, this does not imply anymore that the base manifold is a coset or a constant curvature manifold.
Indeed, let us introduce the covariant derivative of the symplectic
section $(f^\Lambda, f^\Lambda_{\bar I}, \bar f^\Lambda ,
\bar f^\Lambda_I)$ with respect to the $U(1)$--K\"ahler connection $\cQ$
and the spin connection $\omega^{IJ}$ of $\cM_{N=2}$:
 \begin{eqnarray}
& \nabla (f^\Lambda, f^\Lambda_{\bar I}, \bar f^\Lambda, \bar f^\Lambda_I) = & \nonumber\\
& d (f^\Lambda, f^\Lambda_{\bar I}, \bar f^\Lambda,\bar f^\Lambda_I) +
(f^\Lambda, f^\Lambda_{\bar J}, \bar f^\Lambda, \bar f^\Lambda_J)\pmatrix{{\rm i} \cQ   & 0 & 0 & 0 \cr
0 & -{\rm i} \cQ \delta^{\bar I} _{\bar J} + \omega^{\bar I}_{\ \bar J}
& 0 & 0 \cr  0 & 0 & -{\rm i} \cQ & 0 \cr 0 & 0 & 0 &{\rm i} \cQ \delta^{ I} _{ J} + \omega^{I}_{\ J} \cr} &
\end{eqnarray}
where:
 \begin{equation}
\cQ = -{{\rm i} \over 2} ( \partial _i \cK d z^i - \bar \partial_{\bar \imath} \cK  d \bar z^{\bar \imath})
\to d \cQ = {\rm i} g_{i\bar\jmath} dz^i \wedge d\bar z ^{\bar\jmath},
\end{equation}
($\cK$ is the K\"ahler potential)
the K\"ahler weight of $(f^\Lambda, \bar f^\Lambda_I)$ and $(\bar
f^\Lambda, f^\Lambda_{\bar I})$ being $p=1$ and $p=-1$ respectively.
Using the same decomposition as in equation (\ref{defgamma}) and eq.s (\ref{nablav}), (\ref{defomega})
we have in the $N=2$ case:
\begin{eqnarray}
  \label{gammaspec}
  \Gamma &=& \pmatrix{\Omega & \bar \cP \cr \cP & \bar \Omega \cr}
  ,\nonumber\\
  \Omega &=& \omega = \pmatrix{{\rm i} p \cQ  & 0 \cr 0 &
  {\rm i} p \cQ \delta^I_J+ \bar \omega^I_{\ J} \cr}
   \end{eqnarray}
For the embedded vielbein $\cP$ we obtain:
\begin{equation}
\cP = - {\rm i} (f^t \nabla h - h^t \nabla f)  = {\rm i} f^t (\cN -
\bar \cN) \nabla f = \pmatrix{0 & \bar P_I \cr \bar P^{\bar J} & \bar
P^{\bar J}_{\ I} \cr}  \label{viel2}
\end{equation}
where $\bar P^{\bar I} $ is the $(0,1)$--form K\"ahlerian vielbein
while $\bar P^{\bar J}_{\ I} \equiv
 \left( f^t (\cN - \bar \cN) \nabla f \right)^{\bar J}_{\ I}$ is a one--form which in general cannot be expressed
in terms of the vielbein $P^I$ and therefore represents a new geometrical quantity on $\cM_{N=2}$.
Note that we get zero in the first entry of equation  (\ref{viel2})
by virtue of the fact that the identity (\ref{specdef1}) implies $f^\Lambda(\cN -
\bar \cN)_{\Lambda\Sigma}f^\Sigma_i =0 $ and that $f^\Lambda$ is covariantly holomorphic.
If $\Omega $ and $\cP $ are considered as data on $\cM_{N=2}$ then we may interpret
$\Gamma =U^{-1} dU$ as an integrable system of differential equations, namely:
\begin{equation}
  \label{picfuc}
  \nabla (V , V_{\bar I} , \bar V ,  \bar V_I ) =  (V , V_{\bar J} , \bar V , \bar V_J )
  \pmatrix{0& 0 & 0 & \bar P_{I} \cr
0 & 0 & \bar P^{\bar J} & \bar P^{\bar J}_{\ I} \cr 0 &  P_{\bar I} & 0 & 0 \cr  P^J & P^J_{\ \bar I} & 0 & 0 \cr}
\end{equation}
where the flat K\"ahler indices $I,\bar I, \cdots $ are raised and lowered with the flat K\"ahler metric
$\eta_{I\bar J}$.
As it is well known, this integrable system describes the Picard--Fuchs equations for the
periods $(V,V_{\bar I}, \bar V,   V_{I})$ of the Calabi--Yau threefold
with solution given by the $2 n$ symplectic vectors
$V \equiv (f^\Lambda, h_\Lambda)$.
The integrability condition $d\Gamma + \Gamma\wedge \Gamma=0$ gives
three constraints on the K\"ahler base manifold:
\begin{eqnarray}
  d( {\rm i} \cQ) + \bar P^I \wedge  P_{I} &=& 0
  \to \partial_{\bar\jmath}\partial_i \cK = P_{I,i} \bar P_{,\bar \jmath}^I=g_{i\bar\jmath}
  \label{spec1} \\
  (d\omega +\omega\wedge\omega)_{\ \bar I}^{\bar J} &=& P_{\bar I}
  \wedge\bar P^{\bar J} -{\rm i} d\cQ
  \delta_{\bar I}^{\bar J}
-\bar P^{\bar J}_{\ L} \wedge  P^L_{\ \bar I}
\label{spec2}\\
\nabla P_{\ \bar I}^{J} &=& 0
\label{spec3}
\end{eqnarray}
Equation (\ref{spec1}) implies that $\cM_{N=2}$ is a K\"ahler--Hodge manifold.
Equation (\ref{spec2}),
written with holomorphic and antiholomorphic curved indices, gives:
\begin{equation}
  \label{curvspec}
  R_{\bar\imath j \bar k l} = g_{\bar\imath l} g_{j \bar k} + g_{\bar k l}
  g_{i \bar\jmath} - P_{\bar\imath\bar k \bar m}
\bar P_{jln} g^{\bar m n}
\end{equation}
which is the usual constraint on the Riemann tensor of the special geometry.
Note that by consistency of the left and right sides of (\ref{spec2}), the
one-form $P_{\bar \imath \bar k}$
must be a $(0,1)$-form, namely $P_{\bar \imath \bar k}=
P_{\bar \imath \bar k \bar l}d\bar z^{\bar l}$.
The further Special Geometry constraints on the three tensor
$\bar P_{ijk}$  are then consequences of
 equation (\ref{spec3}), which implies:
\begin{eqnarray}
  \nabla_{[l} \bar P_{i]jk} &=& 0 \nonumber\\
\nabla_{\bar l} \bar P_{ijk}&=&0
\label{cijk}
\end{eqnarray}
In particular, the first of equations (\ref{cijk}) also implies
that $ \bar P_{ijk} $ is a completely symmetric tensor.
\par
In summary, we have seen that the $N=2$ theory and the higher
$N$ theories have essentially the same symplectic structure,
the only difference being that since the scalar manifold of $N=2$
is not in general a coset manifold
the symplectic structure allows the presence of a new geometrical
quantity which physically corresponds to
the anomalous magnetic moments of the $N=2$ theory.
It goes without saying that, when $\cM_{N=2}$ is itself a coset manifold \cite{cp},
then the anomalous magnetic moments $ \bar P_{ijk}$
must be expressible in terms of the vielbein of $G/H$.
We give here two examples.
\begin{itemize}
\item{
Suppose $\cM_{N=2}= {SU(1,1) \over U(1)} \times {O(2,n)\over O(2) \times O(n)} $.
The symplectic sections entering the matrix $U$ can be written as follows:
\begin{eqnarray}
  f&=&{\rm i}e^{\cK\over 2}(L^ \Lambda , L^\Lambda_a)\nonumber\\
 h&=&{\rm i}e^{\cK\over 2}(SL^\Sigma\eta_{\Lambda\Sigma},\bar S L^{\Sigma }_a\eta_{\Lambda\Sigma})
\end{eqnarray}
where $\Lambda = 1,\cdots,2n$, $a= 3,\cdots,n$, $\eta_{\Lambda\Sigma}=(1,1,-1,\cdots,-1)$ and we have set
$L^\Lambda = {1\over \sqrt{2}} (L^\Lambda_1 + {\rm i} L^ \Lambda_2)$. In particular from the pseudoorthogonality of
$O(2,n)$ we have:
\begin{equation}
L^\Lambda L^\Sigma \eta_{\Lambda\Sigma} = 0
\end{equation}
Furthermore we have parametrized  ${SU(1,1)\over U(1)}$ as follows:
\begin{equation}
M(S)={1\over {\sqrt{{4 Im S \over 1+\vert S \vert ^2 + 2 Im S}}}}
\left(\matrix{\bfone & {{{\rm i} -S} \over {{\rm i} +S}} \cr
{{{\rm i} + \bar S} \over {{\rm i}- \bar S}} & \bfone \cr }\right)
\end{equation}
We can then compute the embedded connection and vielbein using (\ref{defgamma}).
In particular we find:
\begin{equation}
 \cP = \pmatrix{0 & P_I \cr P_I &- \bar P \delta_{ab} \cr}
\end{equation}
We see that the general $P_{I\bar J}$ matrix in this case can be expressed in terms
of the vielbein of $G/H$ and one finds that the only non vanishing anomalous magnetic moments are:
\begin{equation}
  \bar P_{ab S} = \delta_{ab}P_{,S}= e^{\cK} \delta_{ab}.
\end{equation}}
\item{As a second example we consider the special coset manifold ${SO^*(12) \over U(6)}$.
\par
Note that this manifold also appears as the scalar manifold of the $D=4$, $N=6$ theory, and we refer the reader to section 4
for notations and parametrization of $G/H$.
\par
The integrable system in this case can be written as follows:
\begin{equation}
  \label{so12}
  \nabla \pmatrix{V \cr V_{AB} \cr \bar V \cr \bar V^{AB}\cr} = \pmatrix{0 & 0 & 0 &{1 \over 2} P_{CD}
  \cr 0 & 0 & P_{AB} & {1 \over 2} \bar P_{ABCD} \cr
0 & {1 \over 2}\bar P^{CD} & 0 & 0 \cr \bar P^{AB} & {1 \over 2} P^{ABCD} & 0 & 0 \cr}
\pmatrix{V \cr V_{CD} \cr \bar V \cr \bar V^{CD}\cr}
\end{equation}
where $P^{ABCD}$ is the K\"ahlerian vielbein $(1,0)$--form ($\bar P_{ABCD}=(P^{ABCD})^*$ is a $(0,1)$--form) and:
\begin{equation}
  P_{AB} = {1 \over 4!}\epsilon_{ABCDEF} P^{CDEF} ; \quad \bar P^{AB} = (P_{AB})^*.
\label{pab}
\end{equation}
Moreover, $f_{AB}$ transforms in the {\bf 15} of $SU(6)$, $f$ is an $SU(6)$ singlet and $ \bar f^{AB} = (f_{AB})^*$.
It follows:
\begin{equation}
  \label{so12ol}
  \nabla_i \pmatrix{V \cr V_{AB} \cr \bar V \cr \bar V^{AB}\cr} =
  \pmatrix{0 & 0 & 0 & {1 \over 2} P_{CD,i} \cr 0 & 0 & P_{AB,i} & 0 \cr
0 & 0 & 0 & 0 \cr 0 & {1 \over 2} P^{ABCD}_{,i} & 0 & 0 \cr}  \pmatrix{V \cr V_{CD} \cr \bar V \cr \bar V^{CD}\cr}
\end{equation}
Hence:
\begin{equation}
  \nabla_i \bar V^{AB} = {1 \over 2} P^{ABCD}_{,i} V_{CD}
\end{equation}
If we set:
\begin{equation}
  V_{CD} = P_{CD}^{\ \ \bar l} V_{\bar l} ; \quad \bar V^{AB} = \bar P^{AB,k}V_k
\end{equation}
where the curved indices $k,\bar l$ are raised and lowered with the K\"ahler metric, one easily obtains:
\begin{equation}
  \nabla_i V_j ={1 \over 2} P_{AB, i}P_{CD, j} P^{ABCD}_{,k}g^{k \bar l} f_{\bar l} =
  {1 \over 4}\epsilon _{ABCDEF}
P_{AB, i}P_{CD, j} P_{EF,k}g^{k \bar l} f_{\bar l}.
\end{equation}
Therefore the anomalous magnetic moment is given, in terms of the vielbein, as:
\begin{equation}
\bar P_{ijk} ={1 \over 4} \epsilon _{ABCDEF}P_{AB, i}P_{CD, j} P_{EF,k}
\end{equation}
The other two equations of the integral system give:
\begin{eqnarray}
  \nabla_i V = {1 \over 2} P_{CD,i} V^{CD} &\to & \nabla_i V = V_i \nonumber\\
\nabla_i V_{AB} = P_{AB,i} \bar V &\to & \nabla_i V_{\bar j} = g_{i \bar j} \bar V
\end{eqnarray}
which are the remaining equations defining $N=2$ Special Geometry.
}
\end{itemize}
To complete the analogy between the $N=2$ theory and the higher $N$ theories in $D=4$,
 we also give for completeness, in the $N=2$ case, the central and matter charges, the differential
 relations among them and the sum rules.
 (Note that in $D=4$ $N=2$   we also have
supersymmetric matter given by hypermultiplets whose scalar manifold is quaternionic.
Since we are concerned mainly with central and
matter charges we will not consider the coupling of the hypermultiplets here).
\par
Let us  note that also in $N=2$ the kinetic matrix ${\cal N} _{\Lambda \Sigma}$ which appears in the
vector kinetic Lagrangian \footnote{ The same normalization for the vector kinetic lagrangian will be used
in section 4 when discussing $D=4$, $N>2$ theories}:
\begin{equation}
{\cal L}_{Kin}^{(vector)} = {\rm i} \bar {\cal N} _{\Lambda \Sigma}
{ F}^{-\Lambda}_{\mu\nu} { F}^{-\Sigma \mu\nu}+ h. c.
\end{equation}
\begin{equation}
 { F}^{\pm\Lambda} = {1\over 2} (\bfone \pm \rm i ^\star){ F}^{\Lambda} \label{defv}\\
\end{equation}
\begin{equation}
 {\cG}^-_{\Lambda}=\bar {\cN}_{\Lambda\Sigma}  F^-_{\Lambda} \label{defgd}\\
\end{equation}
is  given in terms of $f, h$ by the
formula:
\begin{equation}
\cN _{\Lambda\Sigma}=h_{\Lambda I}(f^{-1})_ \Sigma ^I \label{nhf}
\end{equation}
The columns of the matrix $f$ appear in the supercovariant
electric field strength $\hat F^\Lambda$:
\begin{equation}
\hat F^\Lambda  = F^\Lambda + f^\Lambda \bar\psi^A \psi ^B
\epsilon_{AB} -{\rm i} \bar f^\Lambda_{\bar\imath}
\bar \lambda^{\bar\imath}_A\gamma_a\psi_B\epsilon_{AB}V^a + h. c.
\end{equation}
(The columns of $h^\Lambda_I$ would appear in the  dual theory
written in terms of the dual magnetic field strengths) .
\par
The transformation laws for the chiral gravitino $\psi_A$ and gaugino
$\lambda^{iA}$ fields are:
\begin{equation}
 \delta \psi_{A \mu}={\cal D}_{\mu}\,\epsilon _A\,+ \epsilon _{AB}
T_{\mu\nu} \gamma^\nu
\epsilon^B + \cdots
 \end{equation}
 \begin{equation}
 \delta\lambda^{iA} = {\rm i} \partial_\mu z^i \gamma^\mu\epsilon^A +
 {{\rm i} \over 2}
T_{\bar \jmath\mu\nu} \gamma^{\mu \nu}
g^{i\bar\jmath}\epsilon^A + \cdots
\end{equation}
where:
\begin{equation}
T \equiv  h_\Lambda F ^{\Lambda}  - f^\Lambda \cG_\Lambda
\end{equation}
\begin{equation}
 T_{\bar \imath} \equiv
 \bar h_{\Lambda_{\bar \imath}} F^{\Lambda}  - \bar f^\Lambda_{\bar \imath} \cG_\Lambda
\end{equation}
are respectively the graviphoton and the matter-vectors
 $z^i$ ($i=1,\cdots,n$) are the complex scalar fields and the position of
 the $SU(2)$ automorphism index A (A,B=1,2) is related to chirality
 (namely $(\psi_A, \lambda^{iA})$ are chiral, $(\psi^A,
 \lambda^{\bar\imath}_A)$ antichiral).
 In principle only the (anti) self dual part of $ F$ and $\cG$ should
 appear in the transformation laws of the (anti)chiral fermi fields; however,
 exactly as in eqs. (\ref{tiden0}),(\ref{tiden}) for $N>2$ theories, from equations (\ref{defgd}), (\ref{nhf}) it follows that :
 \begin{equation}
T^+ = h_\Lambda  F^{+\Lambda} - f^\Lambda \cG_\Lambda ^+   =0
\end{equation}
so that $T=T^-$ (and $\bar T = \bar T^+$).
 Note that both the graviphoton and the matter vectors are $Usp
 (n,n)$ invariant according to the fact that the fermions do not
 transform under the duality group (except for a possible R-symmetry
 phase).
To define the physical charges let us note that in presence of
electric and magnetic sources we can write:
\begin{equation}
\int_{S^2}  F^\Lambda = g^\Lambda , \quad
\int_{S^2} \cG _ \Lambda = e_ \Lambda
\end{equation}
The central charges and the matter charges are now defined as the integrals
over a $S^2$ of the physical graviphoton and matter vectors:
\begin{equation}
Z= \int_{S^2} T= \int_{S^2} ( h_\Lambda  F ^{\Lambda} - f^\Lambda \cG_\Lambda )
= ( h_\Lambda (z,\bar z) g^{\Lambda} - f^\Lambda(z,\bar z) e_\Lambda )
\end{equation}
where $z^i, \bar z^{\bar \imath}$ denote the v.e.v. of the moduli fields in a given
background.
Owing to eq (\ref{dv}) we get immediately:
\begin{equation}
Z_i= \nabla_i Z
\label{fundeq}
\end{equation}
We observe that if in a given background $Z_i =0$ the BPS states in
this configuration have a minimum mass. Indeed
\begin{equation}
\nabla_i Z =0 \rightarrow \partial _i \vert Z \vert ^2 =0.
\end{equation}
 As a consequence of the symplectic structure, one can derive
  two sum
 rules for  $Z$ and $Z_i$:
  \begin{equation}
  \vert Z \vert ^2 \pm  \vert Z_i \vert ^2 \equiv
  \vert Z \vert ^2 \pm   Z_i g^{i\bar \jmath} \bar Z_{\bar \jmath} =
  -{1\over 2} P^t \cM _\pm P
  \label{sumrules}
\end{equation}
  where:
  \begin{equation}
\cM_+ = \left( \matrix{ \bfone & 0 \cr - Re \cN &\bfone\cr}\right)
\left( \matrix{ Im \cN & 0 \cr 0 &Im \cN^{-1}\cr}\right)
\left( \matrix{ \bfone & - Re \cN \cr 0 & \bfone \cr}\right)
\label{m+2}
\end{equation}
\begin{equation}
\cM_- = \left( \matrix{ \bfone & 0 \cr - Re F &\bfone\cr}\right)
\left( \matrix{ Im F & 0 \cr 0 &Im F^{-1}\cr}\right)
\left( \matrix{ \bfone & - Re F \cr 0 & \bfone \cr}\right)
\label{m-2}
\end{equation}
and:
\begin{equation}
P=\left(g^\Lambda, e_ \Lambda \right)
\label{eg2}
\end{equation}
Equation (\ref{m+2}) is obtained by using exactly the same procedure as in (\ref{m+}).
 The sum rule (\ref{m-2}) involves a matrix $\cM_-$, which  has exactly the
same form as $\cM_+$ provided we perform the substitution $\cN_{\Lambda\Sigma}
\rightarrow F_{\Lambda\Sigma}={\partial^2 F \over {\partial X^\Lambda
\partial X^\Sigma}}$($X^\Lambda = e^{-{K\over 2}} f^\Lambda$).
It can be derived in an analogous way by observing
that, when a prepotential $F=F(X)$ exists, Special Geometry gives
the following
extra identity:
\begin{equation}
f^\Lambda_I(F-\bar F)_{\Lambda\Sigma}f^\Sigma_J
= - {\rm i}\eta_{IJ}
\quad \quad \eta =\left(\matrix{ -1 & 0 \cr 0 &  \bfone_{n\times
n}\cr}\right)
\end{equation}
from which it follows:
 \begin{eqnarray}
f \eta f^\dagger &= &{\rm i} \left( F - \bar F \right)^{-1} \\
h \eta h^\dagger &= &{\rm i} \left(\bar F^{-1} - F^{-1} \right)^{-1}\equiv
{\rm i}\bar F \left( F - \bar F \right) ^{-1} F
\end{eqnarray}
Note that while $Im \cN$ has a definite (negative) signature, $Im F$
is not positive definite.

\section{$N>2$ four dimensional supergravities revisited}
\setcounter{equation}{0}
In this section and in the following ones we apply the general considerations of section 3
to the various ungauged supergravity theories with scalar manifold $G/H$ for any $D$ and $N$.
This excludes $D=4$ $N=2$, already discussed in section $3.2$, and $D=5$ $N=2$
for which we refer to the literature.
Our aim is to write down the group theoretical structure of each theory,
 their symplectic or orthogonal  embedding, the vector kinetic
matrix, the supersymmetric transformation laws, the structure
 of the central and matter charges, the differential relations
originating from the Maurer--Cartan equations and the sum rules they satisfy.
For each theory we give the group--theoretical assignments for the fields,
their supersymmetry transformation laws, the ($p+2$)--forms kinetic metrics
and the relations satisfied by central and matter charges.
As far as the boson transformation rules are concerned we prefer to write
 down the supercovariant definition of the
field strengths (denoted by a superscript hat), from which the susy--laws are immediately retrieved.
 As it has been mentioned in section 3 it is here that
the symplectic section $( f^\Lambda _{AB},f^\Lambda_I,\bar f^\Lambda_{AB},
\bar f^\Lambda_I )$ appear as coefficients of the bilinear fermions
in the supercovariant field--strengths while the analogous symplectic
 section $(h_{\Lambda AB},h_{\Lambda I },\bar h_{\Lambda AB},
\bar h_{\Lambda I})$ would appear in the dual magnetic theory.
We include in the supercovariant field--strengths also the supercovariant
vielbein of the $G/H$ manifolds.
Again this is equivalent to giving the susy transformation laws of the scalar fields.
The dressed field strengths from which the central and matter charges
 are constructed appear instead in the susy transformation laws of the
 fermions for which we give the expression up to trilinear fermion terms.
It should be stressed that the numerical coefficients in the aforementioned
 susy transformations and supercovariant field strengths
are fixed by supersymmetry (or ,equivalently, by Bianchi identities in superspace ),
but we have not worked out the relevant computations
being interested in the general structure rather that
 in the precise numerical expressions. However the numerical factors could
also be retrieved by comparing our formulae with those written
 in the standard literature on supergravity and performing the necessary
redefinitions.
The same kind of considerations apply to the central and matter
 charges whose precise normalization has not been fixed.
In the  Tables of  the present  and of the following sections, we give the group assignements for
 the supergravity fields; in particular, we quote the representation
$R_H$ under which the scalar fields of the linearized theory (or the vielbein of $G/H$ of the full theory)
transform.
Furthermore, in  general only the left--handed fermions (when they exist) are quoted.
Right handed fermions transform in the complex conjugate representation of $H$.
In the present section we apply the cosiderations given in section 3.2
to the 4D--Supergravities for $N>2$.
Throughout the section we denote by $A,B,\cdots$ indices of $SU(N)$,
 $SU(N)\otimes U(1)$ being the automorphism group of the $N$--extended
supersymmetry algebra. Lower and upper $SU(N)$ indices on the fermion fields are related to
 their left or right chirality respectively.
If some fermion is a $SU(N)$ singlet chirality is denoted by the usual (L) or (R) suffixes.
Right--handed fermions of $D=4$ transform in the complex f representation of $SU(N)\times U(1) \times H_{matter}$.

Furthermore for any boson field $v$ carrying $SU(N)$ indices
 we have that lower and upper indices are related by complex conjugation,
namely:
\begin{equation}
v_{AB\cdots} = \bar v^{AB\cdots}
\end{equation}
\begin{itemize}
\item{Let us first consider the $N=3$ case \cite{maina}.
The coset space is:
\begin{equation}
G/H= {SU(3,n)\over {SU(3)\otimes SU(n)\otimes U(1)}}
\end{equation}
and the field content is given by:
\begin{eqnarray}
&(V^a_\mu, \psi_{A \mu}, A_\mu , \chi_{(L)}) \quad\quad  A=1,2,3 \quad\quad
\hbox{(gravitational multiplet)}& \\
&(A_\mu , \lambda_A, \lambda_{(R)} , L(z,\bar z))^I  \quad \quad I=1,\cdots,n \quad \quad
\hbox{(vector multiplets)}&
\end{eqnarray}
The transformation properties of the fields are given in the
following Table \ref{4,3} \footnote{We recall that $R_H$  denotes
 the representation which  the vielbein of the scalar
manifold belongs to.}
\begin{table}[ht]
\caption{Transformation properties of fields in $D=4$, $N=3$}
\label{4,3}
\begin{center}
\begin{tabular}{|c||c|c|c|c|c|c|c|c|c|}
\hline
& $V^a_\mu$ & $\psi_{A\mu}$ & $A^\Lambda_\mu$ & $\chi_{(L)}$
& $\lambda^{I}_A$ &
$\lambda^I_{(L)}$ &$ L^\Lambda_{AB} $&$ L^\Lambda_I $& $R_H$ \\
\hline
\hline
$SU(3,n)$ & 1  & 1 & $3+n$ & 1 & 1 & 1 & $3+n$& $3+n$ & -  \\
\hline
$SU(3)$ & 1 & 3 & 1 & 1 & 3 & 1 & 3 & 1 & $3$  \\
\hline
$SU(n)$ & 1  & 1 & 1 & 1 & $n$ & $n$ & 1 & $n$ & $n$ \\
\hline
  $U(1)$ & 0  & ${n\over 2}$ & 0 & 3${n\over 2}$ & 3+${n\over 2}$
  &$-3(1  + {n\over 2})$ & $n$ & $-3$ &$3+n$ \\
\hline
\end{tabular}
\end{center}
\end{table}
The embedding of $SU(3,n)$ in $Usp(3+n,3+n)$ allows to express
the section $(f,h)$ in terms of $L$ as follows::
\begin{eqnarray}
f^\Lambda_{\ \Sigma}&\equiv& (L^\Lambda_{\ AB},\bar L^\Lambda_{\ I})
\label{n3f}  \\
h_{\Lambda \Sigma}&=&{\rm i}(J fJ)_{\Lambda\Sigma} \quad \quad \quad \quad
J= \left(\matrix{\bfone _{3\times 3}&0\cr 0 &-\bfone _{n\times n} \cr} \right) \label{n3h}
\end{eqnarray}
where $AB$ are antisymmetric $SU(3) $ indices, $I$ is an index of
$SU(n)\otimes U(1)$ and $\bar L^\Lambda_{\ I}$ denotes the complex conjugate
of the coset representative.
Using eq.s (\ref{n3f}) and (\ref{n3h}) we have:
\begin{equation}
 \cN_{\Lambda\Sigma} = (h f^{-1})_{\Lambda \Sigma}
 = {1 \over 2} L _\Lambda^{ AB} L_{AB \Sigma} + L_{\Lambda I} \bar L_{I\Sigma} \label{kin2}
\end{equation}
The supercovariant field-strengths and the supercovariant scalar vielbein
 $\hat P^A_I = (L^{-1} \nabla ^{(H)}L)_I^{\ A}$ are:
\begin{eqnarray}
\hat F^\Lambda &=& dA^\Lambda - {1\over 2} f^{\Lambda }_{AB}
 \bar \psi ^A \psi^B + {{\rm i}\over 2} f^{\Lambda }_I
\bar \lambda^I_A \gamma_a \psi^A V^a + {\rm i}f^\Lambda_{AB}\bar
\chi_{(R)}\gamma_a \psi_C \epsilon^{ABC} V^a \nonumber\\
&+& h.c.\\
\hat  P^{\ A}_I &=& P^{\ A}_{I} - \bar\lambda^I_B \psi_C
 \epsilon^{ABC} - \bar \lambda_{I(R)}\psi^A
\end{eqnarray}
where:
\begin{eqnarray}
  P^{I A} & = & {1 \over 2}  \epsilon^{ABC}P_{I BC} =    {1 \over 2}
  \epsilon^{ABC}(L^{-1} \nabla ^{(SU(3) \times U(1))}L)_{I BC}\nonumber\\
&=& P^{IA}_{,i}dz^i \\
\bar P^{IA}& =& P_{IA}
\end{eqnarray}
$z^i$ being the (complex) coordinates of $G/H$
 and $H=H_{Aut} = SU(3) \times U(1)$ .
The chiral fermions transformation laws are given by:
\begin{eqnarray}
\delta \psi_A &=& \nabla \epsilon_A + 2{\rm i} T^{-}_{AB \vert
ab}\Delta^{abc} V_c \epsilon^B + \cdots \\
\delta \chi_{(L)} &=& 1/2 T^{-} _{AB \vert ab} \gamma^{ab} \epsilon_C
\epsilon^{ABC} + \cdots \\
\delta\lambda^I_A &=& -{\rm i} P_{ ,i}^{I B}\partial_a z^i\gamma^a \epsilon^C \epsilon_{ABC}
+ T_{I \vert ab} \gamma^{ab} \epsilon_A + \cdots
\\
 \delta\lambda^I_{(L)} &=& {\rm i} P_{,i}^{I A} \partial_a z^i\gamma^a \epsilon_A
+ \cdots
\end{eqnarray}
 where $T_{AB} $ and $T_{I}$ have the general form given
 in equation (\ref{gravi}).
  Therefore, the general form of the dyonic charges $(Z_{AB}, Z_I)$ are given by eqns.
  (\ref{zab})-- (\ref{charges}).
From the general form of the Maurer-Cartan equations for the embedded
coset representatives $U \in Usp(n,n)$, we find:
\begin{equation}
\nabla^{(H)} \pmatrix{f^\Lambda_{AB} \cr
h_{\Lambda AB}} = \pmatrix{\bar f^\Lambda_I \cr \bar h_{\Lambda I}}
P_I^{\ C} \epsilon_{ABC}
\end{equation}
According to the discussion given in section 3,
using (\ref{zab}), (\ref{zi}) one finds:
 \begin{eqnarray}
\nabla^{(H)} Z_{AB} &=& \bar Z^I
P_I^{\ C} \epsilon_{ABC}   \\
\nabla^{(H)} Z_{I} &=&{1\over 2} \bar Z^{AB}
P_I^{\ C} \epsilon_{ABC}
\end{eqnarray}
and the sum rule:
\begin{equation}
      {1  \over 2} Z ^{  AB} Z_{AB} + Z_{ I} \bar Z_{I} = -{1 \over 2}P^t\cM(\cN)P
\end{equation}
where the matrix $\cM(\cN)$ has the same form as in equation (\ref{m+}) in terms of the kinetic
matrix $ \cN$ of eq.(\ref{kin2}) and $P$ is the charge vector $P^t=(g,e)$.
\item{For $N=4$ \cite{bekose}, the coset space is a  product:
\begin{equation}
G/H= {SU(1,1)\over U(1)}\otimes {O(6,n)\over {O(6) \otimes O(n)}}
\end{equation}
The field content is given by:
Gravitational multiplet:
\begin{equation}
(V^a_\mu,\psi_{A\mu},A_\mu^{AB},\chi_{ABC},S) \quad\quad (A,B=1 ,\cdots,4)
\end{equation}
Vector multiplets:
\begin{equation}
(A_\mu,\lambda^{A},6 \phi)^I \quad \quad (I=1,\cdots,n)
\end{equation}
The coset representative can be written as:
\begin{equation}
L^\Lambda _{\ \Sigma}\to M(S) L^\Lambda_{\ \Sigma}
\end{equation}
where $L^\Lambda_{\ \Sigma}$ parametrizes the coset manifold
 ${O(6,n)\over {O(6)\otimes O(n)}}$
and
\begin{equation}
M(S)={1\over {\sqrt{{4 Im S \over 1+\vert S \vert ^2 + 2 Im S}}}}
\left(\matrix{\bfone & {{{\rm i} -S} \over {{\rm i} +S}} \cr
{{{\rm i} + \bar S} \over {{\rm i}- \bar S}} & \bfone \cr }\right)
\end{equation}
The group assignments of the fields are given in Table \ref{tab4,4}.
\begin{table}[ht]
\caption{$D=4$, $N=4$ transformation properties}
\label{tab4,4}
\begin{center}
\begin{tabular}{|c||c|c|c|c|c|c|c|c|}
\hline
& $V^a_\mu$  & $\psi_{A \vert \mu}$ & $A^\Lambda_{\mu}$ & $\chi_{ABC}$ &
$\lambda_{IA} $ & $ M(S)L^\Lambda_{AB} $ & $M(S) L^\Lambda_I$ & $R_H$ \\
\hline
\hline
$SU(1,1)$ & 1 & 1 & - & 1 & 1 & $2\times 1$ &  $2\times 1$ & - \\
\hline
$O(6,n)$ & 1 & 1 & $6+n$ & 1 & 1 & $1 \times (6+n)$ & $1 \times (6+n)$ & - \\
\hline
$O(6)$ & 1 & 4 & 1 & $\bar 4 $ & $ \bar 4$ & $ 1\times 6$ & 1 & 6 \\
\hline
$O(n)$ & 1 & 1 & 1 & 1 & $n$ & 1 & $n$ & $n$ \\
\hline
$U(1)$& 1 & ${1\over 2}$ & 1 & ${3\over 2}$ & $ -{1\over 2}$ & 1 & 1 & 0 \\
 \hline
\end{tabular}
\end{center}
\end{table}
With the given coset parametrizations the symplectic embedded
 section $(f^\Lambda_\Sigma, h_{\Lambda \Sigma})$ is
(apart from a unessential phase ${{\rm i}+S \over {\rm i} - \bar S} $):
\begin{eqnarray}
f^\Lambda_{\ \Sigma}&=&{\rm i} e^{{K\over 2}}(L^\Lambda_{\ AB},L^\Lambda_{\ I})\\
h_{\Lambda\Sigma}&=&{\rm i} e^{{K\over 2}}(S L^\Gamma_{\ AB}
\eta_{\Lambda\Gamma},\bar S
L^\Gamma_{\ I}\eta_{\Lambda\Gamma})
\end{eqnarray}
where $K=- \mbox{
log}[{\rm i}(S-\bar S)]$ is the K\" ahler potential of ${SU(1,1)\over U(1)}$, and the kinetic matrix
$\cN = h f^{-1}$ takes the form:
\begin{equation}
\cN_{\Lambda\Sigma} = {1\over 2}
(S - \bar S)\bar L_\Lambda^{\ AB} L_{\Sigma AB} +
\bar S \eta _{\Lambda\Sigma}
\label{n44}
\end{equation}
The supercovariant field strengths and the vielbein of the coset manifold are:
\begin{eqnarray}
\hat F^\Lambda &=& dA^\Lambda +\bigl [
f^\Lambda_{AB}(c_1 \bar \psi^A \psi ^B + c_2 \bar \psi_C \gamma_a
\chi^{ABC}V^a)\nonumber\\
&+& f ^\Lambda_I (c_3 \bar\psi^A
\gamma_a \lambda^I_A V^a + c_4 \bar \chi^{ABC} \gamma_{ab} \lambda^{ID}\epsilon_{ABCD}V^a V^b) + h.c. \bigr ]\\
\hat P &=&  P - \bar \psi^A \chi^{BCD} \epsilon_{ABCD}\\
\hat P^I_{AB}&=& P^I_{AB} - (\bar\psi_A \lambda^I_B +\epsilon_{ABCD} \bar\psi^C\lambda^{ID})\\
\end{eqnarray}
where $P= P_{,S} dS$ and $P^I_{AB} =P^I_{AB, i}d\phi^i $ are the vielbein of ${SU(1,1)\over U(1)}$ and ${O(6,n) \over O(6)
     \times O(n)}$
respectively.
The fermion transformation laws are:
\begin{eqnarray}
\delta\psi_A &=& \nabla \epsilon_A +a_1 T^-_{AB \vert ab}\Delta^{abc} \epsilon^B V_c
 + \cdots\\
\delta\chi_{ABC} &=&a_2 P_{,S}\partial_a S \gamma^a \epsilon^D \epsilon_{ABCD} +
a_3 T^-_{[AB \vert ab}\gamma^{ab} \epsilon_{C]} + \cdots\\
\delta\lambda^I_A &=&a_3 P^I_{AB,i}\partial_a \phi^i \gamma^a \epsilon^B + a_4 T^{- I}_{ ab} \gamma^{ab}\epsilon_A + \cdots
\end{eqnarray}
where the 2--forms $T_{AB}$ and $T_{I}$ are defined in eq.(\ref{gravi})
By integration of these two-forms, using eq.(\ref{tiden})--(\ref{charges}) we find the  central and matter dyonic charges
given in eq.s (\ref{zab}), (\ref{zi}).
From the Maurer-Cartan equations for $f,h$ and the
definitions of the charges one easily finds:
\begin{eqnarray}
  \label{charge4,4}
  \nabla^{ SU(4)\otimes U(1)} Z_{AB} &=& \bar  Z^I P_{I AB} + {1 \over 2} \epsilon_{ABCD}\bar Z^{CD}P  \\
  \nabla^{ SO(n)} Z_{I} &=&{1 \over 2} \bar  Z^{AB} P_{I AB} + Z_I
  \bar P
\end{eqnarray}
In terms of the kinetic matrix (\ref{n44}) the sum rule for the charges is given by eqs.(\ref{sumrule})--(\ref{eg}):
 \begin{equation}
 {1 \over 2} Z_{AB} \bar Z^{AB} + Z_I \bar Z_I = -{1\over 2} P^t \cM (\cN) P
\end{equation}
}
}\end{itemize}
For $N>4$ the only available supermultiplet is the gravitational one,
so that $H_{matter}=\bfone$.
The embedding procedure is much simpler than in the matter coupled supergravities since for each $N>4$ there
exists a representation of the scalar manifold isometry group $G$ given in terms of $Usp(n_v,n_v)$ matrices.
\begin{itemize}
\item{ For the $N=5$ theory \cite{dwni}
 the coset manifold is:
\begin{equation}
G/H = {SU(1,5) \over U(5) }
\end{equation}
The field content and the group assignments are displayed in table
\ref{tab4,5}.
\begin{table}[ht]
\caption{Transformation properties of fields in $D=4$, $N=5$}
\label{tab4,5}
\begin{center}
\begin{tabular}{|c||c|c|c|c|c|c|}
\hline
& $V^a$ &$ \psi _A; $ &$\chi_{ABC},\chi_L $
&$A^{\Lambda\Sigma}$ &$L^{x}_{A  }$ & $R_H$  \\
\hline
\hline
$SU(1,5)$& 1 & 1 & 1 & -          &6   & -     \\
\hline
$SU(5)$ & 1 & 5 & 10 + 1 & 1 & 5 &${\bar 5}$ \\
\hline
$U(1)$ & 0 & ${1\over 2}$ & $({3\over 2}, - {5\over 2}
)$ & 0 & 1 & 2 \\
\hline
\end{tabular}
\end{center}
\end{table}
\noindent
Here $x,y,\cdots = 1,\cdots,6$ and  $A,B,C\cdots
=1,\cdots,5$ are  indices of the fundamental representations of $SU(1,5)$ and
 $ SU(5)$,respectively.
$L^x_A$ denote as usual the coset representative in the
fundamental representation of $SU(1,5)$. The antisymmetric couple $\Lambda
\Sigma$, $\Lambda,
\Sigma = 1,\cdots,5$, enumerates the ten vectors.
 The embedding
of $SU(1,5)$ into  the Gaillard-Zumino  group $Usp(10,10)$ is given in terms
of the three-times antisymmetric representation of $SU(1,5)$, a generic element
$t^{xyz}$ satisfying:
\begin{equation}
  \label{txyz}
  t^{xyz}={1 \over 3!} \epsilon^{xyzuvw}t_{uvw}
\end{equation}
 We may decompose $t^{xyz}$ as follows:
\begin{equation}
  \label{txyz1}
  t^{xyz}=\pmatrix{t^{\Lambda\Sigma 6}\cr
t^{\Lambda\Sigma\Gamma}= \epsilon^{\Lambda\Sigma\Gamma \Delta\Pi 6}
t_{\Delta\Pi 6}} \qquad (\Lambda,\Sigma,\cdots= 1,\cdots,5)
\end{equation}
In the following we write $t^{\Lambda\Sigma 6}\equiv t^{\Lambda\Sigma}$.
The 20 dimensional vector $(F^{\mp\Lambda\Sigma}, \cG^\mp_{\Lambda\Sigma})$
transforms under $Sp(20,\IR)$, while, for fixed $AB$, each of the 20-- dimensional vectors
$(f^{\Lambda\Sigma}_{AB},h_{\Lambda\Sigma AB})$ of the
embedding matrix:
\begin{equation}
U = {1\over \sqrt{2}}\pmatrix{f + {\rm i } h & \bar f + {\rm i }\bar h \cr
f - {\rm i } h & \bar f - {\rm i }\bar h \cr}
\end{equation}
 transforms under
$Usp(10,10)$.

The supercovariant field-strengths and  vielbein are:
\begin{eqnarray}
\hat F^{\Lambda\Sigma} &=& d A ^{\Lambda\Sigma} +
\bigl (f^{\Lambda\Sigma}_{\ \ \ AB} (a_1 \bar\psi^A \psi^B + a_2
\bar \psi_C \gamma_a \chi^{ABC}V^a ) + h. c.\bigr ) \\
\hat P_{ABCD} &=& P_{ABCD}- \bar \chi_{[ABC}\psi_{D]}
- \epsilon_{ABCDE} \bar\chi^{(R)} \psi^E
\end{eqnarray}
where $P_{ABCD}= \epsilon_{ABCDF} P^F$ is the complex vielbein, completely antisymmetric
in $SU(5)$
indices and $(P_{ABCD})^{\star} = \bar P^{ABCD}$.\\
The fermion transformation laws are:
\begin{eqnarray}
\delta\psi_A &=& \nabla \epsilon _A  + a_3 T^-_{
 AB \vert ab} \Delta^{abc} \epsilon^B V_c
+ \cdots\\
\delta\chi_{ABC} &=&a_4 P_{ABCD , i}\partial_a \phi^i \gamma^a \epsilon^D +
a_5 T^-_{
 [AB \vert ab}  \gamma^{ab}
\epsilon_{C]} + \cdots \\
\delta \chi_{(L)} &=& a_6 \bar P^{ABCD}_{,\bar i}\partial_a \phi^i \gamma^a \epsilon^E
\epsilon_{ABCDE} + \cdots
\end{eqnarray}
where:
\begin{eqnarray}
  T_{AB}& =&-{{\rm i} \over 2}(\bar f^{-1})_{\Lambda\Sigma AB}F^{\Lambda\Sigma} =
{1 \over 4}(\cN -\bar \cN)_{\Lambda\Sigma,\Gamma\Delta}f^{\Gamma\Delta}_{\ \ AB}
F^{\Lambda\Sigma}\nonumber\\& =&{1 \over 2} (h_{\Lambda\Sigma AB}F^{\Lambda\Sigma} -
f^{\Lambda\Sigma}_{\ \ AB} \cG_{\Lambda\Sigma})\\
\cN_{\Lambda\Sigma, \Delta\Pi}& =& {1 \over 2}
h_{\Lambda\Sigma\vert AB} (f^{-1})^{AB}_{\ \ \Delta\Pi}\\
\cG_{\Lambda\Sigma} &=& -{\rm i}/2{\partial \cL \over \partial F^{\Lambda\Sigma}}
\end{eqnarray}
With a by now familiar procedure one finds the following (complex)
central charges:
\begin{equation}
Z_{AB} = {1 \over 2}( h_{\Lambda\Sigma\vert AB} g^{\Lambda\Sigma}
- f^{\Lambda\Sigma}_{\ \ \ AB}e_{\Lambda\Sigma}
 )
\end{equation}
where:
\begin{eqnarray}
g^{\Lambda\Sigma} &=& \int_{S^2} F^{\Lambda\Sigma}\\
 e_{\Lambda\Sigma} &=& \int_{S^2} \cG_{\Lambda\Sigma}
\end{eqnarray}
From the Maurer--Cartan equation
\begin{equation}
\nabla^{(U(5))} f_{\Lambda\Sigma\vert AB} = {1 \over 2}\bar
f_{\Lambda\Sigma}^{CD}P_{ABCD}
\end{equation}
and the analogous one for $h$ we find:
\begin{equation}
\nabla^{(U(5))} Z_{AB} = {1 \over 2}\bar
Z^{CD}P_{ABCD}
\end{equation}
Finally, the sum rule for the central charges is:
\begin{equation}
{1 \over 2} Z_{AB}\bar Z^{AB} = - {1 \over 2} (g^{\Lambda\Sigma},e_{\Lambda\Sigma})
\cM(\cN)_{\Lambda\Sigma , \Gamma\Delta} \pmatrix{g^{\Gamma\Delta}\cr
e_{\Gamma\Delta}}
\end{equation}
where the matrix $\cM (\cN)$ has exactly the same form as in eq
(\ref{m+}).
}
\item{The scalar manifold of the $N=6$ theory has the coset structure:
\begin{equation}
G/H = {SO^\star (12) \over U(6)}
\end{equation}
We recall that $SO^\star (2n)$ is defined as the subgroup of
$O(2n,\IC)$ that preserves the sesquilinear antisymmetric metric:
\begin{equation}
L^\dagger C L = C, \qquad C= \pmatrix{0 & \bfone \cr -\bfone & 0 \cr}
\end{equation}
The field content and transformation properties are given in
Table \ref{tab4,6},
\begin{table}[ht]
\caption{Transformation properties of fields in $D=4$, $N=6$}
\label{tab4,6}
\begin{center}
\begin{tabular}{|c||c|c|c|c|c|c|}
\hline
& $V^a$ &$ \psi _A$ &$\chi_{ABC},\chi_A$
&$A^\Lambda$ &$S^\alpha_r $  & $R_H$ \\
\hline
\hline
$SO^\star(12)$& 1 & 1 & 1 &- & $ \underline{32}$ & - \\
\hline
$SU(6)$ & 1 & $6$ & $(20 + 6)$ & 1 &$( 15, 1)+(\bar{15},\bar 1)$ & $ {\bar 15}$  \\
\hline
$U(1)$ & 0 & ${1\over 2}$ & $({3\over 2}, -{5\over 2})$ & 0 &$(1,-3) + (-1,3)$ & 2 \\
\hline
\end{tabular}
\end{center}
\end{table}
where $A, B, C = 1,\cdots,6$ are $SU(6)$ indices in the fundamental
representation and $ \Lambda = 1,\cdots, 16 $.
 As it happens  in the $N=5$ theory,
 the $\underline{32}$ spinor
representation of $SO^\star (12)$  can be given in terms of
a  $Usp(16,16)$ matrix, which we denote by $S^\alpha_r$($\alpha,r = 1,\cdots,32 $), so that the embedding
is automatically realized in terms of the spinor representation.
Employing the usual notation we may set:
 \begin{equation}
S^\alpha_r = {1\over \sqrt{2}}\pmatrix{f^\Lambda _I + {\rm i} h_{\Lambda I} &
 \bar f^\Lambda _I + {\rm i} \bar h_{\Lambda I}    \cr
 f^\Lambda _I - {\rm i} h_{\Lambda I} &
 \bar f^\Lambda _I - {\rm i} \bar h_{\Lambda I}    \cr}
\end{equation}
where $\Lambda,I=1,\cdots,16$.
With respect to  $SU(6)$,the sixteen symplectic vectors
$(f^\Lambda_I,h_{\Lambda I})$,
($I = 1,\cdots,16$) are  reducible into the antisymmetric 15--
dimensional representation plus a singlet of $SU(6)$:
\begin{equation}
(f^\Lambda_I,h_{\Lambda I})\to (f^\Lambda_{AB},h_{\Lambda AB})+
(f^\Lambda,h_{\Lambda})
\end{equation}
It is precisely the existence of a $SU(6)$ singlet which allows
for the Special Geometry structure of ${SO^*(12) \over U(6)}$
as discussed in section $3.2$
Note that the coset element $S^\alpha_r$ has no definite $U(1)$
weight since the submatrices
$f^\Lambda_{AB},f^\Lambda$ have the weights 1 and -3 respectively.
The supercovariant field-strenghts and the coset manifold vielbein have the
following expression:
\begin{eqnarray}
\hat F^{\Lambda} &=& d A ^{\Lambda} +\bigl [
f^{\Lambda}_{\  AB} (a_1 \bar\psi^A \psi^B + a_2
\bar \psi_C \gamma_a \chi^{ABC}V^a ) \nonumber\\
&+& a_3
f^\Lambda \bar \psi_C \gamma_a \chi^{C}V^a + h. c. \bigr ] \\
\hat P_{ABCD} &=& P_{ABCD} -  \bar \chi_{[ABC}\psi_{D]}
- \epsilon_{ABCDEF} \bar\chi^E \psi^F
\end{eqnarray}
where $P_{ABCD}=P_{ABCD,i} dz^i$ is the K\"ahler vielbein of the coset.
The fermion transformation laws are:
\begin{eqnarray}
\delta\psi_A &=& \nabla \epsilon _A  + b_1T^-_{AB\vert ab}
 \Delta^{abc} \epsilon^B V_c
+ \cdots\\
\delta\chi_{ABC} &=&  b_2 P_{ABCD\vert a} \gamma^a \epsilon^D +
b_3T^-_{[AB\vert ab} \gamma^{ab}
\epsilon_{C]} + \cdots\\
\delta \chi_A &=& b_4 P^{BCDE\vert a} \gamma^a \epsilon^F
\epsilon_{ABCDEF} +b_5T^-_{ab}\gamma^{ab} \epsilon_A + \cdots
\end{eqnarray}
where:
\begin{eqnarray}
T_{AB}&=&- {\rm i}  (\bar f^{-1})_{\Lambda AB} F^{- \Lambda}\\
T&=& - {\rm i}(\bar f^{-1}) _{\Lambda
} F^{-\Lambda}
\end{eqnarray}
With the usual procedure we have the following complex dyonic central
charges:
\begin{eqnarray}
Z_{AB} &=&  h_{\Lambda AB} g^\Lambda - f^\Lambda_{AB} e_\Lambda \\
 Z &=&   h_{\Lambda} g^\Lambda - f^\Lambda e_\Lambda
\end{eqnarray}
 in the $\underline{15}$ and singlet representation of $SU(6)$
 respectively.
Notice that although we have 16 graviphotons, only 15 central charges
are present in the supersymmetry algebra.
The singlet charge plays a role analogous to a ``matter'' charge.
From the Maurer--Cartan equations:
\begin{eqnarray}
\nabla f^\Lambda_{\ AB} &=&{1 \over 2} \bar f^{\Lambda \vert CD} P_{ABCD} +
{1 \over 4!}\bar f^{\Lambda } \epsilon_{ABCDEF} P^{CDEF} \\
\nabla f^\Lambda &=& {1 \over 2! 4!}f^{\Lambda \vert AB}\epsilon_{ABCDEF}
P^{CDEF}
\end{eqnarray}
and the relation (\ref{nfh-1}) one finds:
\begin{eqnarray}
\nabla ^{(U(6))} Z_{AB} &=& {1 \over 2} \bar Z^{CD} P_{ABCD} + {1 \over 4!}\bar Z
\epsilon_{ABCDEF}P^{CDEF} \\
\nabla^{(U(1))} Z &=& {1 \over 2! 4!}\bar Z^{AB} \epsilon_{ABCDEF}P^{CDEF}
\end{eqnarray}
and the sum-rule (\ref{sumrule}):
\begin{equation}
  \label{sumrule4,6}
{1 \over 2}Z_{AB}\bar Z^{AB} + Z\bar Z = -{1 \over 2} \pmatrix{g^\Lambda,& e_\Lambda \cr}
\cM(\cN)_{\Lambda\Sigma}\pmatrix{
g^\Sigma \cr e_\Sigma \cr}
\end{equation}
with the usual meaning for $\cM(\cN)$ (see eq.(\ref{m+})).
}
\item{In  the $N=8$ case \cite{crju}  the coset manifold is:
\begin{equation}
G/H={E_{7(-7)}\over SU(8)}.
\end{equation}
The field content and group assignments are given in the following Table
\ref{tab4,8}:
\begin{table}[ht]
\begin{center}
  \caption{ Field content and group assignments in $D=4$, $N=8$ supergravity}
  \label{tab4,8}
  \begin{tabular}{|c||c|c|c|c|c|c|}
\hline
&$V^a_\mu $ &$ \psi_A $ & $A^{\Lambda\Sigma}_\mu$ &
 $ \chi_{ABC}$ & $S^\alpha_r$ & $R_H$ \\
\hline
\hline
$E_{7(-7)}$ & 1 & 1 & - & 1 & 56 & - \\
\hline
$SU(8)$ & 1 & 8 & 1 & 56 & $ 28 + \bar{28}$ & 70 \\
\hline
 \end{tabular}
\end{center}
\end{table}

As in $N=5,6$, the embedding is automatically realized in terms
 of the $\underline{56}$ defining representation for $E_7$
 which belongs to $Usp(28,28)$ and it is given by the usual coset element (\ref{defu})
where
\begin{eqnarray}
f+ {\rm i} h& \equiv &f^{\Lambda\Sigma}_{\ \ AB} + {\rm i}
 h_{\Lambda\Sigma AB}\\
\bar f - {\rm i} \bar h& \equiv & \bar f^{\Lambda\Sigma AB} - {\rm i}
 \bar h_{\Lambda\Sigma}^{\ \ AB}
\end{eqnarray}
$\Lambda\Sigma,AB$ are couples of antisymmetric  indices, with
 $\Lambda,\Sigma,A,B$
 running
 from 1 to 8 .
The supercovariant field-strengths and  coset manifold vielbein are:
\begin{eqnarray}
  \hat F^{\Lambda\Sigma} &=& dA^{\Lambda\Sigma} +[ f^{\Lambda\Sigma}_{\ \ AB}(
a_1\bar\psi^A \psi^B + a_2 \bar\chi^{ABC} \gamma_a \psi_C V^a) + h.c.]\\
\hat P_{ABCD} &=&  P_{ABCD} -   \bar\chi_{[ABC}\psi_{D]} + h.c.
\end{eqnarray}
where $ P_{ABCD}= {1 \over 4!} \epsilon_{ABCDEFGH}\bar P^{EFGH}\equiv (L^{-1} \nabla^{SU(8)} L)_{AB\vert CD}=
P_{ABCD,i}d\phi^i$
($\phi^i$ coordinates of $G/H$).
 The fermion transformation laws are given by:
\begin{eqnarray}
  \delta\psi_A &=& \nabla \epsilon_A + a_3 T^-_{AB \vert ab} \Delta^{abc}
\epsilon^B V_c + \cdots\\
\delta\chi_{ABC} &=& a_4 P_{ABCD,a}\gamma^a\epsilon^D + a_5T^-_{[AB\vert ab}
\gamma^{ab} \epsilon_{C]}+ \cdots
\end{eqnarray}
where:
\begin{eqnarray}
  T_{AB}& =&-{{\rm i} \over 2}(\bar f^{-1})_{\Lambda\Sigma AB} F^{\Lambda\Sigma}=
{1 \over 4} (\cN - \bar \cN)_{\Lambda\Sigma, \Gamma\Delta} f^{\Lambda\Sigma} _{AB}
F^{\Gamma\Delta}\nonumber\\
&=& {1 \over 2}(h_{\Lambda\Sigma AB} F^{\Lambda\Sigma}- f^{\Lambda\Sigma}_{\ \ AB}
\cG_{\Lambda\Sigma} )
\end{eqnarray}
with
\begin{eqnarray}
\cN_{\Lambda\Sigma, \Gamma\Delta}&=& {1 \over 2}h_{\Lambda\Sigma AB}
 (f^{-1})^{ AB}_{\ \ \Gamma\Delta}  \\
\cG_{\Lambda\Sigma} &=& -{\rm i/2}{\partial \cL \over \partial F^{\Lambda\Sigma}}
\end{eqnarray}
With the usual manipulations we obtain the central charges:
\begin{equation}
  Z_{AB}={1 \over 2}( h_{\Lambda\Sigma AB}
g^{\Lambda\Sigma} - f^{\Lambda\Sigma}_{\ \ AB} e_{\Lambda\Sigma}),
\end{equation}
the differential relations:
\begin{equation}
  \nabla^{SU(8)}Z_{\ AB}= {1 \over 2} \bar Z^{\ CD} P_{ABCD}
\end{equation}
and the sum rule:
\begin{equation}
 {1 \over 2} Z_{AB}Z^{AB} = - {1 \over 8} (g^{\Lambda\Sigma}, e_{\Lambda\Sigma})
\cM(\cN)_{\Lambda\Sigma , \Gamma\Delta}\pmatrix{g^{\Gamma\Delta} \cr e_{\Gamma\Delta}}
\end{equation}
}
\end{itemize}
\section{Application of the previous formalism to extremal black-holes in four dimensions}
Recently, considerable progress has been made in the study of general
properties
of black holes arising in supersymmetric theories of gravity such as  extended
supergravities, string theory and M-theory \cite{string}.
Of particular interest are extremal black holes in four dimensions which
correspond to BPS saturated states \cite{black} and whose ADM mass depends,
 beyond the
quantized values of electric and magnetic charges, on the asymptotic value
 of scalars at infinity.
The latter describe the moduli space of the theory.
Another physical relevant quantity, which depends only on quantized electric
 and magnetic charges,
 is the black hole entropy,
which can be defined macroscopically, through the Bekenstein-Hawking
 area-entropy relation
or microscopically, through D-branes techniques \cite{dbr} by counting
of microstates \cite{micros}.
It has been further realized that the scalar fields, independently of
their values
at infinity, flow towards the black hole horizon to a fixed value of pure
 topological
nature given by a certain ratio of electric and magnetic charges \cite{fks}.
These ``fixed scalars'' correspond to the extrema of the ADM mass
in moduli space while the black-hole entropy  is the actual value of the
 squared
ADM mass at this point \cite{feka1}.
In theories with $N>2$, extremal black-holes preserving one supersymmetry
have
the further property that all central charge eigenvalues other than the one
equal to the BPS mass flow to zero for ``fixed scalars''.
The black-hole entropy is still given by the square of the ADM mass for
 ``fixed scalars''\cite{feka2}.
Recently \cite{fegika}, the nature of these extrema has been further studied
and shown that  they
generically correspond to non degenerate minima for $N=2$ theories whose
 relevant
moduli space is the special geometry of $N=2$ vector multiplets.
The entropy formula turns out to be in all cases a U-duality invariant
expression
(homogeneous of degree two) built out of electric and magnetic charges and as
 such
can be in fact also computed through certain (moduli-independent) topological
quantities which only depend on the nature of the U-duality groups and the
appropriate representations
of electric and magnetic charges.
For example, in the $N=8$ theory  the entropy was shown to correspond  to the
unique quartic $E_7$ invariant built with its 56 dimensional representation
\cite{kall}.

\section{Central charges, U-invariants and entropy}
In $D=4$, extremal   black-holes   preserving   one   supersymmetry  correspond  to
$N$-extended multiplets with
\begin{equation}
M_{ADM} = \vert Z_1 \vert >  \vert Z_2 \vert  \cdots > \vert Z_{[N/2]} \vert
         \end{equation}
         where $Z_\alpha$, $\alpha =1,\cdots, [N/2]$, are the proper values of
the central charge antisymmetric matrix written in normal form
 \cite{fesazu}.
The central charges $Z_{AB}= -Z_{BA}$, $A,B=1,\cdots,N$, and matter charges
 $Z_I$, $I= 1,\cdots , n$ are
those (moduli-dependent) symplectic invariant combinations of field strenghts
and their duals
(integrated over a large two-sphere)
 which appear
in the gravitino and gaugino supersymmetry variations respectively
\cite{cedafe}, \cite{noi1}, \cite{noi}.
Note that the total number of vector fields is $n_v=N(N-1)/2+n$ (with the
 exception of $N=6$
in which case there is an extra singlet graviphoton)\cite{cj}.
         \\
  It was shown in ref. \cite{feka2} that at the attractor point, where
  $M_{ADM}$ is extremized, supersymmetry requires that $Z_\alpha$, $\alpha >1$,
  vanish together with the matter charges $Z_I$, $I= 1, \cdots , n$
($n$ is the number of matter multiplets, which can exist only for $N=3,4$)
\par
This result can be used to show that for ``fixed scalars'',
corresponding to the attractor point, the scalar ``potential'' of the geodesic
 action \cite{bmgk}\cite{fegika}
\begin{equation}
V=-{1\over 2}P^t\cM(\cN)P
\label{sumrule1}
\end{equation}
is extremized in moduli space.
Here, we recall that $P$ is the symplectic vector $P=(p^\Lambda, q_\Lambda) $ of quantized
electric and
magnetic charges introduced in eqn. (\ref{eg}) and $\cM(\cN)$ is the  $2n_v \times 2n_v$
symplectic matrix
 given in eqn. (\ref{m+}).

The above assertion  is the result of computing the extremum of (\ref{sumrule1})  by use of equations
(\ref{dz1}), (\ref{sumrule}). We obtain:
\begin{equation}
\label{zz}
  P^{ABCD} Z_{AB} Z_{CD} =0 ;\quad Z_I=0
\end{equation}
$P_{ABCD}$ being the vielbein of the scalar manifold, completely antisymmetric
in its $SU(N)$ indices.
It is easy to see that in the normal frame these equations imply:
\begin{eqnarray}
  \label{norm}
  M_{ADM}\vert _{fix} & \equiv & \vert Z_{1} \vert \neq 0 \\
 \vert Z_{i} \vert & =& 0 \qquad (i=2, \cdots ,N/2 )
\end{eqnarray}

The main purpose of this section is to provide  particular expressions which
 give the
entropy formula as a moduli-independent quantity in the entire
moduli space and not just at the critical points.
Namely, we are looking for quantities $S\left(Z_{AB}(\phi), \bar Z^{AB}
 (\phi),Z_{I}(\phi), \bar Z^{I} (\phi)\right)$
such that ${\partial \over \partial \phi ^i} S =0$, $\phi ^i$ being the moduli
 coordinates.

These formulae generalize the quartic $E_{7(-7)}$ invariant of $N=8$
supergravity \cite{kall} to all other cases.
\par
Let us first consider the theories $N=3,4$, where  matter can be
present \cite{maina}, \cite{bks}.
\par
The U-duality groups
 are, in these cases, $SU(3,n)$ and $SU(1,1)
\times SO(6,n)$ respectively (Here we denote by U-duality group the isometry
 group $G$
acting on the scalars, although only a restriction of it to integers is the
 proper U-duality group \cite{ht}).
The central and matter charges $Z_{AB}, Z_I$ transform in an obvious
way under the isotropy groups
\begin{eqnarray}
H&=& SU(3) \times SU(n) \times U(1) \qquad (N=3) \\
 H&=& SU(4) \times O(n) \times U(1) \qquad (N=4)
\end{eqnarray}
Under the action of the elements of $G/H$ the charges get mixed with
their complex conjugate.
\\
For $N=3$:
\begin{eqnarray}
P^{ABCD}&=& P_{IJ}=0 \, , \, \, P_{ AB I}  \equiv  \epsilon_{ABC}P^C_I \nonumber\\
\quad Z_{AB}
 &\equiv & \epsilon_{ABC}Z^C
 \label{viel3}
\end{eqnarray}
Then  the variations are:
\begin{eqnarray}
\delta Z^A  &=& \xi^A _I \bar Z^I  \\
\delta Z_{I}  &=&\xi^A _{ I} \bar Z_A
\label{deltaz3}
\end{eqnarray}
where $\xi^A_I$ are infinitesimal parameters of $K=G/H$.  Indeed,
once the covariant derivatives are known, the variations are obtained
by the  substitution $\nabla \to \delta$, $P \to \xi$.
\par
With a simple calculation, the U-invariant expression is:
\begin{equation}
S=   Z^A \bar Z_A - Z_I \bar Z^I
\label{invar3}
\end{equation}
In other words, $\nabla_i S = \partial_i S =0 $, where the covariant
derivative is
defined in ref. \cite{noi}.
\par
Note that at the attractor point ($Z_I =0$) it coincides with the
moduli-dependent potential (\ref{sumrule1})
computed at its extremum.
\\
For $N=4$
\begin{eqnarray}
P_{ABCD} &=& \epsilon_{ABCD}P ,\quad P_{IJ} = \eta_{IJ}\bar
P \nonumber\\
P_{AB I}&=& {1\over 2} \eta_{IJ} \epsilon_{ABCD}\bar P^{CD J} \label{viel4}
\end{eqnarray}
and the transformations of $K= {SU(1,1) \over U(1)} \times
{O(6,n) \over O(6) \times O(n)}$ are:
 \begin{eqnarray}
\delta Z_{AB}  &=& {1\over 2} \xi  \epsilon_{ABCD} \bar Z^{CD}  +
 \xi _{AB I} \bar Z^I \\
\delta Z_{I}  &=& \bar \xi \eta_{IJ} \bar Z^J + {1\over 2}\xi _{AB I} \bar Z^{AB}
\label{deltaz4}
\end{eqnarray}
with $\bar \xi^{AB I} =  {1\over 2} \eta^{IJ}   \epsilon^{ABCD}
\xi_{CD J}$.
\par
There are three $O(6,n)$ invariants given by $I_1$, $I_2$, $\bar I_2$ where:
\begin{eqnarray}
  I_1 &=& {1 \over 2}  Z_{AB} \bar Z_{AB} - Z_I \bar Z^I
\label{invar41} \\
I_2 &=& {1\over 4} \epsilon^{ABCD}  Z_{AB}   Z_{CD} - \bar Z_I \bar Z^I
\label{invar42}
\end{eqnarray}
and the unique   $SU(1,1)  \times
O(6,n) $  invariant $S$, $\nabla S =0$, is given by:
\begin{equation}
S= \sqrt{(I_1)^2 - \vert I_2 \vert ^2 }
\label{invar4}
\end{equation}
At the attractor point $Z_I =0$ and $\epsilon^{ABCD} Z_{AB} Z_{CD}
=0$ so that $S$ reduces to the square of the BPS mass.
\par
For $N=5,6,8$ the U-duality invariant expression $S$ is the square
root of a unique invariant under the corresponding U-duality groups
$SU(5,1)$, $O^*(12)$ and $E_{7(-7)}$.
The strategy is to find a quartic expression $S^2$    in terms of
$Z_{AB}$ such that $\nabla S=0$, i.e. $S$ is moduli-independent.
\par
As before, this quantity is a particular combination of the $H$
quartic invariants.
\par
For $SU(5,1)$ there are only two  $U(5)$ quartic invariants.
In terms of the matrix $A_A^{\ B} = Z_{AC} \bar Z^{CB}$ they are:
$(Tr A)^2$, $Tr(A^2)$, where
\begin{eqnarray}
 Tr A & = & Z_{AB} \bar Z^{BA} \\
 Tr (A^2) & = & Z_{AB} \bar Z^{BC} Z_{CD} \bar Z^{DA}
\end{eqnarray}
As before, the relative coefficient is fixed by the transformation
properties of $Z_{AB}$ under ${SU(5,1) \over U(5) } $ elements of
infinitesimal parameter $\xi^C$:
\begin{eqnarray}
  \delta Z_{AB} = {1\over 2} \xi^C \epsilon_{CABPQ} \bar Z^{PQ}
\end{eqnarray}
It then follows that the required invariant is:
\begin{equation}
S= {1\over 2} \sqrt{  4 Tr(A^2) - (Tr A)^2 }
\label{invar5}
\end{equation}
For $N=8$ the $SU(8)$ invariants are:
\begin{eqnarray}
I_1 &=& (Tr A) ^2 \\
I_2 &=& Tr (A^2) \\
I_3 &=& Pf \, Z   \nonumber\\
 &=& {1\over 2^4 4!} \epsilon^{ABCDEFGH} Z_{AB} Z_{CD} Z_{EF} Z_{GH}
\end{eqnarray}
The ${E_{7(-7)} \over SU(8)}$ transformations are:
\begin{equation}
\delta Z_{AB} ={1\over 2} \xi_{ABCD} \bar Z^{CD} \label{transf8}
\end{equation}
where $\xi_{ABCD}$ satisfies the reality constraint:
\begin{equation}
\xi_{ABCD} = {1 \over 24} \epsilon_{ABCDEFGH} \bar \xi^{EFGH}
\end{equation}
One finds the following $E_{7(-7)}$ invariant \cite{kall}:
\begin{equation}
S= {1\over 2} \sqrt{4 Tr (A^2) - ( Tr A)^2 + 32 Re (Pf \,
Z) }
\end{equation}
The $N=6$ case is the more complicated because under $U(6)$ the
left-handed spinor of $O^*(12)$ splits into:
\begin{equation}
32_L \to (15,1) + ( \bar {15}, -1) + (1, -3) + (1,3)
\end{equation}
The transformations of ${O^*(12) \over U(6)}$ are:
\begin{eqnarray}
\delta Z_{AB} &=& {1\over 4} \epsilon_{ABCDEF} \xi^{CD} \bar Z^{EF} +
\xi_{AB} \bar X \nonumber\\
\delta X &=& {1 \over 2} \xi _{AB} \bar Z^{AB} \label{transf6}
\end{eqnarray}
 where we denote by $X$ the $SU(6)$ singlet.
The quartic $U(6)$ invariants are:
\begin{eqnarray}
I_1&=& (Tr A)^2 \label{invar61}\\
I_2&=& Tr(A^2)\label{invar62} \\
I_3 &=& Re (Pf \, ZX) \nonumber\\
&=& {1\over 2^3 3!}
Re( \epsilon^{ABCDEF}Z_{AB}Z_{CD}Z_{EF}X)\label{invar63}\\
I_4 &=& (Tr A) X \bar X\label{invar64}\\
I_5&=& X^2 \bar X^2\label{invar65}
\end{eqnarray}
The unique $O^*(12)$ invariant is:
\begin{eqnarray}
S&=&{1\over 2} \sqrt{4 I_2 - I_1 + 32 I_3 +4I_4 + 4 I_5 }
\label{invar6}   \\
\nabla S &=& 0
\end{eqnarray}
Note that at the attractor point $Pf\,Z =0$, $X=0$ and $S$ reduces to
the square of the BPS mass.

 We note that most of the results given above can be obtained in a
 simple way by performing the relevant computations in the so called
 ``normal frame'' of the $Z_{AB}$ matrix.

 Indeed, in order to determine the quartic U-invariant expressions $S^2$ , $\nabla S
 =0$, of the $N>4$ theories, it is useful to use, as a calculational tool,
transformations
 of the coset which preserve the normal form of the $Z_{AB}$ matrix.
 It turns out that these transformations are certain Cartan elements
 in $K=G/H$ \cite{solv}, that is they belong to $O(1,1)^p\in K$, with $p=1$
for $N=5$,
 $p=3$ for $N=6,8$.
 \par
 These elements act only on the $Z_{AB}$ (in normal form), but they
 uniquely determine the U-invariants since they mix the eigenvalues
 $e_i$ ($i=1,\cdots,[N/2]$).
 \par
 For $N=5$, $SU(5,1)/U(5)$ has rank one (see ref. \cite{gilmore}) and
 the element is:
 \begin{equation}
\delta e_1 = \xi e_2 ; \quad \delta e_2 = \xi e_1
\end{equation}
which is indeed a $O(1,1)$ transformation with unique invariant
\begin{equation}
\vert (e_1)^2 - (e_2) ^2\vert = {1 \over 2} \sqrt{ 8 \left((e_1)^4 + (e_2)^4
\right) - 4 \left( (e_1)^2 + (e_2)^2 \right)^2 }
\end{equation}
For $N=6$, we have $\xi_1 \equiv \xi_{12}; \xi_2 \equiv \xi_{34};
\xi_3 \equiv \xi_{56}$ and we obtain the 3 Cartan elements of
$O^*(12) /U(6)$, which has rank 3, that is it is a $O(1,1)^3$ in $O^*
(12)/U(6)$.
Denoting by $e$ the singlet charge, we have the following $O(1,1)^3$
transformations:
\begin{eqnarray}
\delta e_1 &=& \xi_2 e_3  + \xi _3 e_2   + \xi _1 e \label{trans61}\\
 \delta e_2 &=& \xi_1 e_3  + \xi _3 e_1   + \xi _2 e\label{trans62} \\
 \delta e_3 &=& \xi_1 e_2  + \xi _2 e_1   + \xi _3 e\label{trans63} \\
 \delta e &=& \xi_1 e_1  + \xi _2 e_2   + \xi _3 e_3\label{trans64}
\end{eqnarray}
these transformations fix uniquely the $O^*(12)$ invariant constructed out of
the five $U(6)$ invariants displayed in (\ref{invar61}-\ref{invar65}).
For $N=8$ the infinitesimal parameter is $\xi_{ABCD}$ and, using the
reality condition, we get again a $O(1,1)^3$ in $E_{7(-7)} /SU(8)$.
Setting $\xi_{1234}= \xi_{5678} \equiv \xi_{12}$, $\xi_{1256}=
\xi_{3478} \equiv \xi_{13}$,
$\xi_{1278}= \xi_{3456} \equiv \xi_{14}$, we have the following set of
transformations:
\begin{eqnarray}
 \delta e_1 &=& \xi_{12} e_2  + \xi _{13} e_3   + \xi _{14} e_4\label{trans81}
 \\
 \delta e_2 &=& \xi_{12} e_1  + \xi _{13} e_4   + \xi _{14} e_3\label{trans82}
 \\
 \delta e_3 &=& \xi_{12} e_4  + \xi _{13} e_1   + \xi _{14} e_2\label{trans83}
 \\
 \delta e_4 &=& \xi_{12} e_3  + \xi _{13} e_2   + \xi _{14} e_1\label{trans84}
\end{eqnarray}
These transformations fix uniquely the relative coefficients of the three
 $SU(8)$
invariants:
\begin{eqnarray}
I_1&=& 4 (e_1^2 + e_2^2 + e_3^2 + e _4^2)^2 \\
I_2&=& 2 (e_1^4 + e_2^4 + e_3^4 + e _4^4) \\
I_3&=& e_1e_2e_3e _4 \\
\end{eqnarray}
It is easy to see that the transformations (\ref{trans61}-\ref{trans64}) and
 (\ref{trans81}-\ref{trans84}) correspond to three commuting matrices
(with square equal
to $\bfone$):
\begin{equation}
\pmatrix{0&0&0&1 \cr 0&0&1&0 \cr 0&1&0&0 \cr 1&0&0&0 \cr} ;
\pmatrix{0&1&0&0 \cr 1&0&0&0 \cr 0&0&0&1 \cr 0&0&1&0 \cr} ;
\pmatrix{0&0&1&0 \cr 0&0&0&1 \cr 1&0&0&0 \cr 0&1&0&0 \cr}
\end{equation}
which are proper non compact Cartan elements of $K$.
The reason we get the same transformations for $N=6$ and $N=8$
is because the extra singlet $e$ of $N=6$ can be identified
with the fourth eigenvalue of the central charge of $N=8$.

\section{Extrema of the BPS mass and fixed scalars}
In this section we would like to extend the analysis of the extrema of the
black-hole induced potential
\begin{equation}
V= {1\over 2} Z_{AB}\bar Z^{AB} + Z_I \bar Z^ I
\end{equation}
which was performed in ref \cite{fegika} for the $N=2$ case to all $N>2$
 theories.
We recall that, in the case of $N=2$ special geometry with metric
 $g_{i\bar\jmath}$,
 at the fixed scalar critical point
$\partial _i V=0$ the Hessian matrix reduces to:
\begin{eqnarray}
( \nabla_i \nabla _{\bar\jmath} V)_{fixed}&=&
( \partial_i \partial _{\bar\jmath} V)_{fixed} = 2g_{i\bar\jmath}V_{fixed}\\
( \nabla_i \nabla _j V)_{fixed}&=&0
\end{eqnarray}
The Hessian matrix is strictly  positive-definite if the
critical point is not at the singular point of the
vector multiplet moduli-space.
This matrix was related to the Weinhold metric earlier  introduced in
the geometric
approach to thermodynamics and used for the study of critical phenomena
\cite{fegika}.
For $N$-extended supersymmetry, a form of this matrix was also given and
 shown to be
equal to \footnote{Generically the indices $i,j$ refer to real coordinates,
unless
the manifold is K\"ahlerian, in which case we use holomorphic coordinates and
 formula (\ref{hessian})
reduces to the hermitean $i\bar\jmath$ entries of the Hessian matrix.}:
\begin{eqnarray}
V_{ij}&=& ( \partial_i \partial _j V)_{fixed} \nonumber\\
&=& Z_{CD} Z^{AB} ({1\over 2}
 P^{CDPQ}_{\ \ \ \ ,j}P_{ABPQ,i} \nonumber\\
 &+&
P^{CD}_{I,j} P^I_{AB,j}).
\label{hessian}
\end{eqnarray}
It is our purpose to further investigate properties of the Weinhold metric
 for fixed scalars.
Let us first observe that the extremum conditions $\nabla_i V=0$, using the
 relation
between the covariant derivatives of the central charges, reduce to the
 conditions:
\begin{equation}
\epsilon^{ABCDL_1 \cdots L_{N-4}}Z_{AB} Z_{CD} = 0,\quad Z_I=0
\label{fixedsc}
\end{equation}
These equations give the fixed scalars in terms of electric and magnetic
charges
and also show that the topological invariants of the previous section reduce
 to the extremum of the square of the ADM mass since, when the above
 conditions are fulfilled,
$(Tr A)^2 = 2Tr(A^2)$, where $A_A^{\ B} = Z_{AB} \bar Z^{BC}$.
On the other hand, when these conditions are fulfilled, it is easy to see
that the Hessian matrix is degenerate.
To see this, it is sufficient to go, making an $H$ transformation, to the
normal frame in which these conditions imply $Z_{12} \neq 0 $ with the other
 charges vanishing.
Then we have:
\begin{eqnarray}
&& \partial_i \partial_j V \vert_{fixed} =\nonumber\\
&& 4 \vert Z_{12}\vert ^2
({1\over 2} P^{12ab} _j P_{12ab,i} +  P^{ 12 I }_{,j} P_{12 I,i}) ,\quad
\label{normetr}
\end{eqnarray}
where $a,b \neq 1,2$.
To understand the pattern of degeneracy for all $N$, we observe that when only
 one central
charge is not vanishing the theory effectively reduces to an $N=2$ theory.
Then the actual degeneracy respects $N=2$ multiplicity of the scalars degrees
of freedom in the
sense that the degenerate directions will correspond to the hypermultiplet
 content
of $N>2$ theories when decomposed with respect to $N=2$ supersymmetry.

Note that for $N=3$, $N=4$, where $P_{AB I} $ is present, the Hessian is
block diagonal.
For $N=3$, referring to eq. (\ref{viel3}), since the  scalar manifold is
K\"ahler, $P_{AB I}$ is a (1,0)-form while $P^{ AB I}= \bar P_{AB I}$
is a (0,1)-form.
\par
The scalars appearing in the
$N=2$ vector multiplet and hypermultiplet content of the vielbein
are $P_{3I}$ for the vector multiplets and $P_{aI}$  ($a=1,2$) for
the hypermultiplets.
 From equation (\ref{normetr}), which for the $N=3$ case reads
 \begin{equation}
  \partial_{\bar \jmath} \partial_i V \vert_{fixed} = 2 \vert Z_{12}\vert ^2
 P_{3I,\bar \jmath} P^{3I}_{,i}
\end{equation}
we see that the metric has $4n$ real directions corresponding to
$n$ hypermultiplets  which are degenerate.
\par
For $N=4$, referring to (\ref{viel4}),
 $P$ is the $SU(1,1)/U(1)$
vielbein which gives one matter vector multiplet scalar while $P_{ 12 I}$ gives
$n$ matter vector multiplets.
The directions which are hypermultiplets correspond to $P_{ 1a I}$, $P_{ 2a I}$
 ($ a=3,4$).
Therefore the ``metric'' $V_{i j}$ is of rank $2n+2$.
\vskip 5mm
\par
For $N>4$, all the scalars are in the gravity multiplet and correspond to
 $P_{ABCD}$.
\par
The splitting in vector and hypermultiplet scalars proceeds as
before.
Namely, in the $N=5$ case we set $P_{ABCD}=\epsilon_{ABCDL}P^L $
($A,B,C,D,L =1,\cdots 5$). In this case the vector multiplet scalars
are $P^a$ ($a=3,4,5$) while the hypermultiplet scalars are $P^1,
P^2$ ($n_V = 3$, $n_h =1$).
\par
For $N=6$, we set $P_{ABCD} ={1\over 2} \epsilon_{ABCDEF} P^{EF}$. The vector
multiplet scalars are now described by $P^{12}, P^{ab} $
($A,B,...=1,...,6$; $a,b = 3,\cdots 6$), while the hypermultiplet
scalars are given in terms of $P^{1a}, P^{2a} $.
Therefore we get $n_V = 6+1 =7$, $n_h =4$.
\par
This case is different from the others because, besides the
hypermultiplets $P^{1a}, P^{2a}$, also the vector multiplet direction
$P^{12}$ is degenerate.
Finally, for $N=8$ we have $P_{1abc}, P_{2abc}  $ as hypermultiplet
scalars and $P_{abcd}$ as vector multiplet scalars, which give
$n_V=15$, $n_h = 10$ (note that in this case the vielbein satisfies a
reality condition: $P_{ABCD}= {1\over 4! } \epsilon_{ABCDPQRS} \bar P
^{PQRS}$).
We have in this case 40  degenerate directions.
\par
In conclusion we see that the rank of the matrix $V_{ij} $ is $(N-2)
(N-3) + 2 n$ for all the four dimensional theories.

\section{Relations to string theories}
$N$-extended supergravities are related to strings compactified on
six-manifolds $M_N$ preserving $N$ supersymmetries at $D=4$.
Since we are presently considering $N>2$, the most common cases are
$N=4$ and $N=8$.
The first can be achieved in heterotic or Type II string, with $M_4 =
T_6$ in heterotic and $M_4 = K_3 \times T_2$ in Type II theory.
These theories are known to be dual at a non perturbative level \cite{ht},
 \cite{witten}, \cite{dlr}.
$N=8$ corresponds to $M_8 = T_6$ in Type II.
\par
Less familiar are the $N=3,5$ and 6 cases which were studied in ref.
\cite{fkff}.
\par
Interestingly enough, the latter cases can  be obtained by
compactification of Type II on asymmetric orbifolds with $3= 2_L +
1_R$, $5 = 4_L + 1_R$ and $6 = 4_L + 2_R$ respectively.
\par
BPS states considered in this paper should correspond to massive
states in these theories for which only a subset of them is known in
the perturbative framework.
\par
In attemps to test non perturbative string properties it would be
interesting to check the existence of the BPS states and their
entropy by using microscopic considerations.
\par
We finally  observe that, unlike $N=8$, the moduli spaces of
$N=3,5,6$ theories are locally K\"ahlerian (as $N=2$) with coset
spaces of rank 3 ($n \geq 3$), 1 and 3 respectively.
\par
For $N=5,6$ these spaces are also special  K\"ahler (which is also
the case for $N=3$ when $n = 1,3$) \cite{crvp} \cite{cfg2}.
\par
We  can use the previous observations to construct U-invariants
for some  $N=2$ special geometries looking at the representation content of
 vectors and their duals
with respect to U-dualities.
Let us first consider $N=2$ theories with U-duality $SU(1,n)$ and $SU(3,3)$.
These groups emerge in discussing string compactifications on some $N=2$
 orbifolds
(i.e. orbifold points of Calabi-Yau threefolds)\cite{cfg2}\cite{seib}.
The vector content is respectively given by the fundamental representation of
 $SU(1,n)$
and the twenty dimentional threefold antisymmetric rep. of $SU(3,3)$
 \cite{ffs}.
Amazingly, the first representation occurs as in $N=3$ matter coupled theories,
while the latter  is the same as in $N=5$ supergravity
(note that $SU(1,n)$,  $SU(3,n)$ and  $SU(3,3)$,  $SU(5,1)$ are just different
non compact forms of the same $SU(m)$ groups).
From the results of the previous section we conclude that
  the special manifolds  ${SU(1,n)\over SU(n)\times U(1)}$
and ${SU(3,3)\over SU(3)\times SU(3)\times U(1)}$
 admit respectively a quadratic \cite{cedafe}, \cite{sabra} and
a quartic topological invariant.
The $N=2$  special manifold ${O^*(12)\over U(6)}$ has a vector content
which is a left spinor of $O^*(12)$, as in the $N=6$ theory, therefore it
 admits
a quartic invariant.
Finally, the $N=2$ special manifolds ${SU(1,1) \over U(1)}
 \times{O(2,n)\over O(2) \times O(n)}$, which emerge in $N=2$
compactifications
of both heterotic and Type II strings \cite{seib}, admit a quartic invariant
 which can be read
from  the $N=4$ quartic invariant in which the  ${SU(1,1) \over U(1)}$ matter
charge is identified
with the second eigenvalue of the $N=4$ central charge.
All the above topological invariants can then be interpreted as entropy
of a variety of $N=2$ black-holes.


\end{document}